\documentstyle[preprint,aps,prd,eqsecnum,tighten,amsfonts,amssymb,epsfig]{revtex} 
\newcommand{\PhiH}{\text{$\Phi_{\text{{\tiny H}}}$}}
\newcommand{\PsiH}{\text{$\Psi_{\text{{\tiny H}}}$}}
\newcommand{\PsibH}{\text{$\bar{\Psi}_{\text{{\tiny H}}}$}}
\newcommand{\vphiH}{\text{$\varphi_{\text{{\tiny H}}}$}}
\newcommand{\psib}{\text{$\bar{\psi}$}}
\newcommand{\yc}{\text{$f$}}
\newcommand{\phih}{\text{$\hat{\phi}$}}
\newcommand{\Mpl}{\text{$M_{\text{{\tiny P}}}$}}
\newcommand{\phihp}{\text{$\hat{\phi}_{+}$}}
\newcommand{\phihm}{\text{$\hat{\phi}_{-}$}}
\newcommand{\phip}{\text{$\phi_{+}$}}
\newcommand{\phim}{\text{$\phi_{-}$}}
\newcommand{\vphip}{\text{$\varphi_{+}$}}
\newcommand{\vphim}{\text{$\varphi_{-}$}}
\newcommand{\psip}{\text{$\psi_{+}$}}
\newcommand{\psim}{\text{$\psi_{-}$}}
\newcommand{\psibp}{\text{$\bar{\psi}_{+}$}}
\newcommand{\psibm}{\text{$\bar{\psi}_{-}$}}
\newcommand{\vphi}{\text{$\varphi$}}

\begin{document}
\draft
\preprint{UMDPP\#97-127}
\title{Nonequilibrium inflaton dynamics and reheating.\\
II. Fermion production, noise, and stochasticity}
\author{S. A. Ramsey,\thanks{Electronic address: {\tt sramsey@physics.umd.edu}}
B. L. Hu,\thanks{Electronic address: {\tt hub@physics.umd.edu}}
and A. M. Stylianopoulos}
\address{Department of Physics, University of Maryland,
College Park, Maryland 20742-4111}
\date{Submitted to Phys.\ Rev. D on August 13, 1997}
\maketitle
\begin{abstract}
We present a detailed and systematic analysis of the coarse-grained,
nonequilibrium dynamics
of a scalar inflaton field coupled to a fermion field in the latter stages
of the reheating period of inflationary cosmology.  We derive coupled 
nonperturbative equations for the inflaton mean field 
and variance at two loops in a general
curved spacetime, and show that the equations of motion are real and causal, 
and that the gap equation for the two-point function is dissipative due to
fermion particle production.
We then specialize to the case of Minkowski space and 
small-amplitude inflaton oscillations, and derive the perturbative one-loop
dissipation and noise kernels to fourth order in the
Yukawa coupling constant; the normal-threshold dissipation
and noise kernels are shown to satisfy a zero-temperature 
fluctuation-dissipation relation.  We derive a Langevin equation for the 
dynamics of the inflaton zero mode, and show that its
variance is non-negligible during reheating.  Stochastic evolution of the
inflaton may have important astrophysical consequences.
\end{abstract}
\newpage

%%%%%%%%%%%%%%%%%%%%%%%%%%%%%%%%%%%%%%%%%%%%%%%%%%%%%%%%%%%%%%%%%%%%%%%%%%%%%%
%                                                                            %
% SECTION I:  Introduction and summary                                       %
%                                                                            %
%%%%%%%%%%%%%%%%%%%%%%%%%%%%%%%%%%%%%%%%%%%%%%%%%%%%%%%%%%%%%%%%%%%%%%%%%%%%%%
\section{Introduction}
In an earlier paper \cite{ramsey:1997b} we studied the effect of parametric
particle creation on the inflaton dynamics in the post-inflation,
``preheating'' stage.  It was shown that analysis of the dominant physical 
processes during the early stages of reheating necessitates consideration
of self-consistent back reaction of the inflaton field variance on the inflaton
mean field and quantum modes.  This is a {\em nonperturbative effect,\/} 
and its description requires a consistent truncation of the Schwinger-Dyson 
equations \cite{calzetta:1995b,boyanovsky:1996b,ramsey:1997b}.  Assuming 
initial conditions
conducive to efficient parametric particle creation, the end state of the
regime of parametric particle creation consists of a large inflaton variance
(i.e., on the order of the tree level terms in the inflaton's effective mass). 
Thus, both the inflaton mean field and variance should be treated on
an equal footing.  This requires a two-particle-irreducible (2PI)
formulation of the effective action which is a subclass of the master
effective action \cite{calzetta:1995b}.

During the later stages of reheating, the dynamics of the inflaton field is
thought, in the case of unbroken symmetry,
to be dominated by damping from fermionic particle 
creation  \cite{kofman:1997a}.  This stage of inflaton dynamics is the subject 
of this paper.  We consider a model consisting of a scalar inflaton field 
$\phi$ (with $\lambda \phi^4$ self-coupling) 
coupled to a spinor field $\psi$ via a Yukawa interaction and  
%largely of a phenomenological nature 
%\cite{abbott:1982a,albrecht:1982a,dolgov:1982a,morikawa:1985a},
we attempt to present as complete and rigorous a treatment as mandated by the 
self-consistency of the formalism and the
actual solvability of the equations. Thus, we adopt a closed-time-path (CTP),
two-particle-irreducible (2PI), coarse-grained effective action (CGEA) to
derive the dynamics of the inflaton field. We have explained the significance
of CTP and 2PI in our earlier papers 
\cite{calzetta:1987a,calzetta:1988b,calzetta:1989a,calzetta:1989b,hu:1989a,calzetta:1995b,ramsey:1997a} and their relevance to the study of inflaton dynamics
in Ref.~\cite{ramsey:1997b}.
Here, an added feature of an {\em open system\/} 
is introduced: we wish to include the averaged effect of an environment
on the system, and a useful method is via the coarse-grained effective action
\cite{hu:1991b,sinha:1991a,lombardo:1996a,dalvit:1996a,greiner:1997a}.
Let us explain the rationale for this.

{\em Coarse-Grained Effective Action:\/}
In inflationary cosmology at the onset of the reheating period, the inflaton 
field's zero mode typically has a large expectation value, 
whereas all other fields coupled to the inflaton, as
well as inflaton modes with momenta greater than the Hubble constant, are 
to a good approximation in a vacuum state 
\cite{brandenberger:1985a}. This suggests imposing a physical
coarse-graining in which one regards the inflaton field as the system, and the 
various quantum fields coupled to it as the environment.  From the
closed-time-path, coarse-grained effective action (CTP-CGEA) 
\cite{hu:1991b,sinha:1991a,lombardo:1996a,dalvit:1996a,greiner:1997a}
derived in Sec.~\ref{sec-necgidcst} below,
one obtains effective dynamical equations for the inflaton field, taking into 
account its effect on the environment, and back reaction therefrom.
For the present problem, the system consists of the inflaton mean field and 
variance, and the environment consists of the spinor field(s) coupled to the 
system via a Yukawa interaction.

We wish to emphasize here a subtle yet important distinction between the 
system-environment division in nonequilibrium statistical mechanics
and the system-bath 
division assumed in thermal field theory.  In the latter, one assumes that 
the propagators for the bath degrees of freedom are {\em fixed,\/}
finite-temperature equilibrium Green functions, whereas in the case of the
CTP-CGEA, the environmental propagators are {\em slaved\/} (in the sense of
\cite{calzetta:1995b}) to the dynamics of
the system degrees of freedom, and are not fixed {\em a priori\/} to be
equilibrium Green functions for all time.  This distinction is important for
discussions of fermion particle production during reheating, because it is 
only when the inflaton mean field amplitude is small enough for the use of
perturbation theory, that the system-bath split implicit in thermal
field theory can be used.  Otherwise,
one must take into account the effect of the inflaton mean field on the
bath (spinor) {\em propagators.\/}  
%The Pauli blocking effect, which, crudely
%speaking, says that the particle production into a particular mode of the
%fermion field with spatial momentum $\vec{k}$ is suppressed by a factor
%$(1-2n_{\vec{k}})$, where $n_{\vec{k}}$ is the occupation number for the
%$\vec{k}$ mode. 
%This effect has been shown analytically only in the 
%former case, i.e., small-amplitude inflaton oscillations and 
%finite-temperature field theory \cite{boyanovsky:1995d}.

{\em Earlier work:\/}
The first studies of particle production during reheating in inflationary
cosmology were \cite{abbott:1982a,dolgov:1982a}, where reasonable estimates
of particle production were made, but back reaction effects were not addressed.
The earliest studies of fermionic particle creation during reheating used
time-dependent perturbation theory to compute the imaginary part of
the self-energy for the zero mode of the inflaton field, which was
related to the damping parameter in a friction-type phenomenological 
term in the equation of motion for the inflaton zero mode
\cite{abbott:1982a,albrecht:1982a,morikawa:1985a,dolgov:1990a,shtanov:1995a}.
In these studies, it was
assumed that the effect of fermionic particle production
on the dynamics of the inflaton zero mode is that of a $\Gamma \dot{\phi}$
friction term.  However, it has been shown 
\cite{morikawa:1986a,calzetta:1989a,paz:1990a,stylianopoulos:1991a,kofman:1994a,boyanovsky:1995a,boyanovsky:1995d,kofman:1996a} 
that this assumption is not tenable for a wide variety of field-theoretic
interactions and initial conditions.  Rather, the effect of back reaction from 
particle production must be accounted for by {\em deriving\/} the effective
evolution equation for the quantum expectation value of the inflaton zero mode,
where the dynamics of the degrees of freedom of the produced particles are 
either coarse-grained (as in Sec.~\ref{sec-dsasfo} of this paper), 
or accounted for through 
self-consistent coupled equations (as in Sec.~\ref{sec-necgidcst}).  In 
general, particle creation leads to a {\em nonlocal}
dissipation term in the inflaton mean field equation, and it is only under
rather idealized conditions and specialized cases that one can expect the
dissipation kernel to approach a delta function (i.e., a friction term)
\cite{hu:1992a,hu:1993c,hu:1993a,hu:1994a}.  Therefore, 
\cite{abbott:1982a,dolgov:1990a,shtanov:1995a} missed
the time-nonlocal nature of fermion particle production and its effect
on the dynamics of the inflaton field.  In addition, these
studies computed the self-energy in flat space, thereby neglecting the
effect of curved spacetime on fermion production, and did not examine
the stochastic noise arising from the coarse-graining of the fermion
field.  

In addition, most early studies of fermion production during reheating, in
using time-dependent perturbation theory to compute the {\em vacuum\/} particle
production rate \cite{abbott:1982a,dolgov:1990a,shtanov:1995a},
did not include the effect of back reaction of the {\em produced\/} fermion 
particles
on the particle production process itself.  In \cite{boyanovsky:1995d},
the effect of a thermal initial fermion distribution on the particle production
process was investigated (and a Pauli blocking effect was shown), but
their analytic derivation of the Pauli blocking effect involves the same
perturbative expansion (i.e., system-bath split) described above,
and therefore does not incorporate the effect of the {\em produced\/} 
fermion quanta on the particle production process.  In order to
take this effect into account, it is necessary to include the effect of the
time-varying inflaton mean field in the equation of motion for the spinor
propagator, which amounts to a coarse-graining of the fermion field, in the
system-environment sense, as described above.  The perturbative amplitude 
expansion of the effective inflaton dynamics, in contrast, amounts to 
a system-bath coarse-graining which does not include this back reaction 
effect.

More recent studies of fermion production during reheating 
\cite{boyanovsky:1995b,boyanovsky:1995c,boyanovsky:1995d}
obtained dynamical equations for the inflaton mean field using
a one-loop factorization of the Lagrangian,
%the one-particle-irreducible (1PI), closed-time-path, coarse-grained
%effective action, 
and solved them numerically.  However, these studies
were carried out in flat space, and because they studied only the dynamics
of the inflaton mean field (and at one loop), their analysis did not
take into account the back reaction of the inflaton variance on the
fermion quantum modes, nor the back reaction of particle production
on the dynamics of the inflaton two-point function.  The importance
of the curved spacetime effect has been addressed in a previous
paper \cite{ramsey:1997b}, and we will discuss below the importance
of treating the inflaton quantum variance on equal footing with the mean field.

{\em Present work:\/} In this study, we wish to 
describe the late stages of the reheating period, in which the damping of
the inflaton mean field is dominated by fermionic particle production;
our starting point is the end of the {\em preheating\/} period (in which
the inflaton dynamics was dominated by back reaction from parametric
particle creation, as discussed in our earlier paper \cite{ramsey:1997b})\@.  
Because the inflaton variance $\langle \varphi^2 \rangle$ can (for
sufficiently strong self-coupling) be on the order 
of the square of the amplitude of mean field oscillations at the end of 
preheating (in the case of unbroken symmetry)
\cite{ramsey:1997b}, it is necessary to treat the inflaton
mean field and variance on an equal footing in a study of the subsequent
effective dynamics of the inflaton field.  This requires a two-loop,
two-particle-irreducible formulation of the coarse-grained 
effective action.  At two loops, both the
inflaton mean field {\em and the inflaton variance\/} couple to the spinor 
degrees of freedom, and are damped by back reaction from fermion 
particle production; all the previous studies mentioned 
above, in using the 1PI effective action, missed this possibly important
effect.  
%
%neglecting the latter, as is the practice in previous 
%studies of inflaton dynamics in the regime of fermion particle production 
%\cite{abbott:1982a,dolgov:1990a,shtanov:1995a} is therefore inconsistent
%because the inflaton variance is on the order of the tree-level effective
%mass.

In addition to having a large variance, 
the inflaton amplitude at the end of the preheating period 
may be large enough that the usual perturbative expansion 
in powers of the Yukawa coupling constant 
is not valid [see Eq.~(\ref{eq-vpt}) below], 
in which case a nonperturbative derivation of the inflaton
dynamics is required.  
%Previous studies of the nonperturbative inflaton dynamics during the stage of
%fermion particle production 
%\cite{boyanovsky:1995b,boyanovsky:1995c,boyanovsky:1995d} 
%relied on the one-loop, one-particle-irreducible (1PI) effective action, 
%which is inadequate.  
In the construction
of the CTP-2PI, coarse-grained effective action below, the spinor
propagators obey one-loop dynamical equations in which the inflaton mean
field acts as an external source.  
Studies of fermion particle production during reheating which relied 
on the use of perturbation theory in the Yukawa coupling constant
\cite{abbott:1982a,dolgov:1982a,albrecht:1982a,morikawa:1985a,dolgov:1990a,shtanov:1995a,kofman:1994a}
therefore do not apply to the case of fermion particle production
at the end of preheating with unbroken symmetry, 
when the Yukawa coupling is sufficiently large.
The dynamical equations derived
in Sec.~\ref{sec-necgidcst}
below for the inflaton mean field and variance are applicable even when, as
may be the case, the inflaton mean field amplitude is large enough that a 
perturbative expansion in powers of the Yukawa coupling is not justified.  

Although, as discussed above, a proper treatment of the early stage of fermion
production during reheating, starting at the end of preheating with
unbroken symmetry, should in principle employ the CTP-2PI-CGEA to obtain
coupled equations of motion for the inflaton mean field and variance, 
at very late times the inflaton mean field and variance will have been
damped sufficiently (due to the dissipative mechanisms derived below in 
Sec.~\ref{sec-necgidcst}) that the perturbative 1PI effective action
will yield a qualitatively correct description of the inflaton mean-field dynamics.
While curved spacetime effects should in principle be incorporated
self-consistently for a quantitative calculation of the reheating
temperature in a particular inflationary model (as discussed in 
\cite{ramsey:1997b}), for a general discussion of dissipative 
effective dynamics of the inflaton mean field in the case of
weak cosmic expansion (where the Hubble constant $H$ is much smaller
than the frequency of inflaton oscillations), it is reasonable to
neglect curved spacetime effects in computing the spinor propagators.
Therefore in Sec.~\ref{sec-dsasfo}, we derive the perturbative, 
flat-space CTP-1PI-CGEA to 
fourth order in the Yukawa coupling constant, and obtain an
evolution equation for the inflaton mean field with nonlocal dissipation.

Another new feature of our work obtainable only from the stochastic approach
adopted here is the derivation, in Sec.~\ref{sec-nk},
of a Langevin equation for the inflaton mean
field, with clear identification of the dissipation and noise kernels from
the CGEA. We have calculated the energy dissipated and the fluctuations
in the energy. From the latter we obtain the range of parameters where
the conventional ``mean-field'' approach breaks down. We believe the
methodology presented here provides a better theoretical 
framework for the investigation of phase transitions in the early universe, 
as exemplified by our treatment of reheating in inflationary cosmology.

{\em Organization and Notation:\/}
Our work is organized as follows.  In Sec.~\ref{sec-necgidcst}, we derive
the coupled equations of motion for the inflaton mean field and variance,
in a general curved spacetime, including diagrams in the coarse-grained CTP-2PI
effective action of up to two-loop order.  In Sec.~\ref{sec-dsasfo},
we specialize to Minkowski space and small mean-field amplitude, and obtain
a perturbative mean field equation including dissipative effects up to
fourth order in the Yukawa coupling constant.  In Sec.~\ref{sec-nk}, we 
examine the dissipation and noise kernels obtained in Sec.~\ref{sec-dsasfo},
and show that they obey a fluctuation-dissipation relation.  We then derive a 
Langevin equation which self-consistently includes the effect of
noise on the dynamics of the inflaton zero mode.  We summarize our results in
Sec.~\ref{sec-conc}.

Throughout this paper we use units in 
which $c = 1$.  Planck's constant $\hbar$ is shown explicitly 
(i.e., not set equal to 1) except in those sections where noted. 
In these units, Newton's constant is $G = \hbar \Mpl^{-2}$, where
$\Mpl$ is the Planck mass.  We assume a four-dimensional, globally
hyperbolic, orientable spacetime manifold, and follow
the sign conventions\footnote{In the classification
scheme of Misner, Thorne and Wheeler \cite{misner:1973a}, the 
sign convention of Birrell and Davies \cite{birrell:1982a} is 
classified as $(+,+,+)$.}  
of Birrell and Davies \cite{birrell:1982a}
for the metric tensor $g_{\mu\nu}$, 
the Riemann curvature tensor $R_{\mu\nu\sigma\rho}$, and
the Einstein tensor $G_{\mu\nu}$.  We use greek letters to 
denote spacetime indices. The beginning latin letters 
$a,b,c,d,e,f$ indicate the time branch (see Ref.\ \cite{ramsey:1997a},
Sec.~II).
The Einstein summation convention over repeated indices is employed.

%%%%%%%%%%%%%%%%%%%%%%%%%%%%%%%%%%%%%%%%%%%%%%%%%%%%%%%%%%%%%%%%%%%%%%%%%%%%%%
%                                                                            %
% SECTION II:  Nonequilibrium dynamics in FRW spacetime                      %
%                                                                            %
%%%%%%%%%%%%%%%%%%%%%%%%%%%%%%%%%%%%%%%%%%%%%%%%%%%%%%%%%%%%%%%%%%%%%%%%%%%%%%
\section{Coarse-grained inflaton dynamics in curved spacetime}
\label{sec-necgidcst}
In this section, we present a first-principles derivation of the 
nonequilibrium, nonperturbative, effective dynamics of a scalar inflaton 
field $\phi$ coupled to a spinor field $\psi$ via a Yukawa interaction, 
in a general, curved classical background spacetime.  The use of the 
Schwinger-Keldysh closed-time-path (CTP) formalism allows formulation
of the nonequilibrium dynamics of the inflaton from an appropriately
defined initial quantum state.
The evolution equations for the inflaton mean field and variance are derived 
from the two-loop, closed-time-path (CTP), two-particle-irreducible (2PI), 
coarse-grained effective action (CGEA).  
As the name suggests, there are
two approximations of a statistical mechanical nature.  One is the 
coarse graining of the environment--- here the inflaton field is the system and
the fermion field is the environment \cite{hu:1991b}.  
The other refers to a 
truncation of the {\em correlation hierarchy\/} for the inflaton field 
\cite{calzetta:1995b}--- the two-particle-irreducible effective action.  This
formulation retains the inflaton mean field and variance as coupled dynamical
degrees of freedom.  
Back Reaction of the scalar and spinor
field dynamics on the spacetime is incorporated using the semiclassical 
Einstein equation, which follows from functional differentiation of the
effective action with respect to the metric.   It is shown that
these dynamical equations are both real and causal, and the
equations are cast in a form suitable for implementation of an explicit
curved-spacetime renormalization procedure.

\subsection{The model}
\label{sec-model}
We study a model consisting of a scalar field $\phi$ (the 
inflaton field) which is Yukawa-coupled to a spinor field $\psi$, in a curved,
dynamical, classical background spacetime.  The total action
\begin{equation}
S[\phi,\bar{\psi},\psi,g^{\mu\nu}] = S^{\text{{\tiny G}}}[g^{\mu\nu}] +
S^{\text{{\tiny F}}}[\phi,\bar{\psi},\psi,g^{\mu\nu}],
\label{eq-ca}
\end{equation}
consists of a part depicting classical gravity, 
$S^{\text{{\tiny G}}}[g^{\mu\nu}]$, and a part for the matter fields,
\begin{equation}
S^{\text{{\tiny F}}}[\phi,\bar{\psi},\psi,g^{\mu\nu}] = 
S^{\phi}[\phi,g^{\mu\nu}] + S^{\psi}[\bar{\psi},\psi,g^{\mu\nu}] +
S^{\text{{\tiny Y}}}[\phi,\bar{\psi},\psi,g^{\mu\nu}], 
\end{equation}
whose scalar (inflaton), spinor (fermion), and Yukawa-interaction 
parts are given by
\begin{mathletters}
\begin{eqnarray}
\label{eq-sfa}
&& S^{\phi}[\phi,g^{\mu\nu}] = - \frac{1}{2} \int d^{\;4}x \sqrt{-g} \left[
\phi ( \square + m^2 + \xi R ) \phi +  \frac{\lambda}{12} \phi^4 \right], \\
&& S^{\psi}[\bar{\psi},\psi,g^{\mu\nu}] = \int d^{\;4}x \sqrt{-g} 
\left[ \frac{i}{2} \left(\bar{\psi} \gamma^{\mu} \nabla_{\mu} \psi -
(\nabla_{\mu} \bar{\psi}) \gamma^{\mu}\psi\right)
 - \mu \bar{\psi}\psi \right], \\
&& S^{\text{{\tiny Y}}}[\phi,\bar{\psi},\psi,g^{\mu\nu}] =
- \yc \int d^{\;4}x \sqrt{-g} \phi \bar{\psi} \psi.
\label{eq-sfc}
\end{eqnarray}
\end{mathletters}
For this theory to be renormalizable in semiclassical gravity,
the bare gravity action $S^{\text{{\tiny G}}}[g^{\mu\nu}]$
of Eq.~(\ref{eq-ca}) should have the form \cite{dewitt:1975a,birrell:1982a}
\begin{equation}
S^{\text{{\tiny G}}}[g^{\mu\nu}] = \frac{1}{16 \pi G} \int d^{\;4}x 
\sqrt{-g} \left[ R - 2 \Lambda_c + c R^2 + b R^{\alpha\beta} R_{\alpha\beta}
+ a R^{\alpha\beta\gamma\delta}R_{\alpha\beta\gamma\delta}\right].
\label{eq-sg}
\end{equation}
In Eqs.~(\ref{eq-sfa})--(\ref{eq-sfc}), $m$ is the scalar field ``mass'' 
(with dimensions of inverse 
length); $\xi$ is the dimensionless coupling to gravity; $\mu$ is the spinor
field ``mass,'' with dimensions of inverse length; $\square$ is the 
Laplace-Beltrami operator in the curved background spacetime with metric
tensor $g_{\mu\nu}$; $\nabla_{\mu}$ is the covariant derivative compatible with
the metric; $\sqrt{-g}$ is the square root of the absolute value of the
determinant of the metric; $\lambda$ is the self-coupling of the inflaton
field, with dimensions of $1/\sqrt{\hbar}$; and 
$\yc$ is the Yukawa coupling constant,
which has dimensions of $1/\sqrt{\hbar}$.  In Eq.~(\ref{eq-sg}), 
$G$ is Newton's constant (with dimensions of length divided by mass);
$R$ is the scalar curvature; $R_{\mu\nu}$ is
the Ricci tensor; $R_{\alpha\beta\gamma\delta}$ is the Riemann tensor;
$a$, $b$, and $c$ are constants with dimensions of length squared; and
$\Lambda_c$ is the cosmological constant, which has dimensions of
inverse length-squared.  The curved spacetime Dirac matrices $\gamma^{\mu}$ 
satisfy the anticommutation relation
\begin{equation}
\left\{ \gamma^{\mu},\gamma^{\nu} \right\}_{+} = 2 g^{\mu\nu} 1_{\text{{\tiny
sp}}},
\end{equation}
in terms of the contravariant metric tensor $g^{\mu\nu}$.  The symbol 
$1_{\text{{\tiny sp}}}$ denotes the identity element in the Dirac algebra.  

Due to the generalized Gauss-Bonnet theorem \cite{chern:1962a}, the
constants $a$, $b$, and $c$ are not all independent in four spacetime
dimensions; let us therefore set $a=0$.  It is assumed that there is a 
definite separation
of time scales between the stage of ``preheating'' discussed in our earlier
work \cite{ramsey:1997b}, and fermionic particle production, which is our
primary focus in this work.  However, this does not imply that perturbation
theory in the Yukawa coupling constant
$f$ is necessarily valid, which would require that condition
(\ref{eq-vpt}) (defined in Sec.~\ref{sec-dsasfo} below) be satisfied.
In addition, the fermion field mass $\mu$ is assumed to be 
much lighter than the inflaton field mass $m$, i.e., the renormalized
parameters $m$ and $\mu$ satisfy $m \gg \mu$.

\subsection{Closed-time-path, coarse-grained effective action}
We denote the quantum Heisenberg field operators of the scalar field $\phi$ 
and the spinor field $\psi$ by $\PhiH$ and $\PsiH$, respectively, and the 
quantum state\footnote{Although in this case the particular initial 
conditions constitute a pure state, this formalism can encompass general 
mixed-state initial conditions \cite{calzetta:1988b}.} by $|\Omega\rangle$.  
For consistency with the truncation of the correlation hierarchy at second 
order, we assume $\Phi_{\text{{\tiny H}}}$ to have a Gaussian moment expansion 
in the position basis \cite{mazzitelli:1989b}, in which case the relevant 
observables are the scalar mean field
\begin{equation}
\hat{\phi}(x) \equiv \langle \Omega | \PhiH(x) | \Omega \rangle,
\end{equation}
and the mean-squared fluctuations, or variance, of the scalar
field
\begin{equation}
\langle \Omega | \Phi^2_{\text{{\tiny H}}} (x)| \Omega \rangle 
- \langle \Omega | \PhiH(x) | \Omega \rangle^2
\equiv \langle \Omega | \vphiH^2(x) | \Omega \rangle,
\label{eq-sv}
\end{equation}
where the last equality follows from the definition of the scalar
{\em fluctuation field\/}
\begin{equation}
\vphiH (x) \equiv \PhiH(x) - \phih(x).
\label{eq-ff}
\end{equation}
As discussed above, at the end of the preheating period, 
the inflaton variance can be as large as the square of
the amplitude of mean-field oscillations.
On the basis of our assumption of separation of time scales in 
Sec.~\ref{sec-model},
and the conditions which prevail at the onset of reheating, the initial 
quantum state $|\Omega\rangle$ is assumed to be an appropriately defined
vacuum state for the {\em spinor\/} field.  

The construction of the CTP-2PI-CGEA for the $\phi\psib\psi$ theory
in a general, curved, background spacetime
closely parallels the construction of the CTP-2PI effective action
for the O($N$) model in curved spacetime discussed in an earlier paper
\cite{ramsey:1997a}.  Within the spacetime manifold (whose dynamics must be
determined self-consistently through the semiclassical gravitational
field equation), let $M$ be defined as the past domain of dependence of a 
Cauchy hypersurface $\Sigma_{\star}$, i.e., $M \equiv D_{-}(\Sigma_{\star}),$
where $\Sigma_{\star}$ has been chosen to be far to the future of any
dynamics we wish to study.
We now define a ``CTP'' manifold ${\mathcal M}$ as the quotient space
\begin{equation}
{\mathcal M} \equiv ( M \times \{+,-\} )/\sim,
\end{equation}
in terms of the discrete set $\{+,-\}$ which labels the time branch.
The volume form $\bbox{\epsilon}_{\mathcal M}$ on ${\mathcal M}$
is defined to be $\bbox{\epsilon}_M$ on the plus ($+$) branch and 
$-\bbox{\epsilon}_M$ on the minus ($-$) branch.  The restrictions of
a function $\phi$, defined on ${\mathcal M}$, to the $+$ and $-$ time branches
are denoted by $\phi_{+}$ and $\phi_{-}$, respectively.  
We can then define a matter field action on ${\mathcal M}$,
\begin{equation}
{\mathcal S}^{\text{{\tiny F}}}[\phi_{-},\psib_{-},\psi_{-},g^{\mu\nu}_{-};
\phi_{+},\psib_{+},\psi_{+},g^{\mu\nu}_{+}] \equiv
S^{\text{{\tiny F}}}[\phi_{+},\psib_{+},\psi_{+},g^{\mu\nu}_{+}] -
S^{\text{{\tiny F}}}[\phi_{-},\psib_{-},\psi_{-},g^{\mu\nu}_{-}],
\label{eq-sfctp}
\end{equation}
where the spacetime integrations in $S^{\text{{\tiny F}}}$ are now over $M$ 
only.  We use the symbol ${\mathcal S}^{\text{{\tiny F}}}$ to distinguish it 
from the action $S^{\text{{\tiny F}}}$ on $M$.  Let us also simplify
notation by suppressing time branch indices in the argument of functionals
on ${\mathcal M}$, i.e.,
\begin{equation}
{\mathcal S}^{\text{{\tiny F}}}[\phi,\psib,\psi,g^{\mu\nu}] \equiv
{\mathcal S}^{\text{{\tiny F}}}[\phi_{-},\psib_{-},\psi_{-},g^{\mu\nu}_{-};
\phi_{+},\psib_{+},\psi_{+},g^{\mu\nu}_{+}].
\end{equation}
Let us also define the functional 
${\mathcal S}^{\text{{\tiny Y}}}$ on ${\mathcal M}$ by
\begin{equation}
{\mathcal S}^{\text{{\tiny Y}}}[\phi,\psib,\psi,g^{\mu\nu}] \equiv
S^{\text{{\tiny Y}}}[\phi_{+},\psib_{+},\psi_{+},g^{\mu\nu}_{+}] -
S^{\text{{\tiny Y}}}[\phi_{-},\psib_{-},\psi_{-},g^{\mu\nu}_{-}],
\label{eq-syctp}
\end{equation}
in analogy with Eq.~(\ref{eq-sfctp}).  
For a function $\phi$ on ${\mathcal M}$, the restrictions of $\phi$ to the
$+$ and $-$ time branches are subject to the boundary condition
\begin{equation}
(\phi_{+})_{|\Sigma_{\star}} = (\phi_{-})_{|\Sigma_{\star}}
\label{eq-bc}
\end{equation}
at the hypersurface $\Sigma_{\star}$.
The gravity action $S^{\text{{\tiny G}}}$, promoted to
a functional on ${\mathcal M}$, takes the form
\begin{equation}
{\mathcal S}^{\text{{\tiny G}}}[g^{\mu\nu}_{+},g^{\mu\nu}_{-}] = 
S^{\text{{\tiny G}}}[
g^{\mu\nu}_{+}] - S^{\text{{\tiny G}}}[g^{\mu\nu}_{-}],
\label{eq-sgm}
\end{equation}
where the range of spacetime integration in $S^{\text{{\tiny G}}}$ on 
the right-hand side of Eq.~(\ref{eq-sgm}) is understood
to be over $M$.  

To formulate the CTP-2PI-CGEA, our first step is to define a generating 
functional for $n$-point functions of the scalar field, in terms of
the initial quantum state $|\Omega\rangle$ which evolves under the 
influence of a local source $J$, and a non-local source $K$ coupled to 
the scalar field (in the interaction picture with the external sources
being treated as the ``interaction'').  This generating
functional depends on both $J$ and $K$, as well as the
classical background metric $g^{\mu\nu}$.  In the path integral representation,
the generating functional $Z[J,K,g^{\mu\nu}]$ takes the form of a sum over 
scalar field configurations $\phi$ and complex Grassmann-valued configurations 
$\psi$ on the manifold ${\mathcal M}$,
\begin{eqnarray}
Z[J,K,g^{\mu\nu}] \equiv 
\int_{\text{{\tiny ctp}}} && D\phi_{-} D\bar{\psi}_{-} D\psi_{-} 
D\phi_{+} D\bar{\psi}_{+} D\psi_{+} \exp \Biggl[ 
\frac{i}{\hbar} \biggl(  {\mathcal S}^{\text{{\tiny F}}}[\phi,\bar{\psi},
\psi,g^{\mu\nu}] \nonumber \\
&& + \int_M d^{\;4}x \sqrt{-g} c^{ab} J_a \phi_b  \nonumber \\ 
&& + \frac{1}{2} \int_M d^{\;4}x \sqrt{-g} \int_M d^{\;4}x' \sqrt{-g'} c^{ab}
c^{cd} K_{ac}(x,x') \phi_b(x) \phi_d(x') \biggr) \Biggr],
\label{eq-zgf}
\end{eqnarray}
where $J_a(x)$ is a local $c$-number source and $K_{ab}(x,x')$ is a nonlocal
$c$-number source.  The subscript CTP on the functional integral denotes a
summation over field configurations $\phi_{\pm}$, $\psib_{\pm}$, and 
$\psi_{\pm}$ which satisfy the boundary condition (\ref{eq-bc}).  The latin 
indices $a$, $b$, $c$, $\ldots\;$, have the discrete index set 
$\{ +,- \}$, and denote the time branch \cite{calzetta:1987a,calzetta:1988b}. 
The boundary conditions on the functional integral of Eq.~(\ref{eq-zgf}) 
at the initial data surface determine the quantum state 
$|\Omega\rangle$. The CTP indices have been dropped from $g^{\mu\nu}$ for ease 
of notation; it will be clear how to reinstate them \cite{ramsey:1997a} 
in the two-loop CTP--2PI effective action shown below in 
Sec.~\ref{sec-dsasfo}.  The two-index symbol $c^{ab}$ is defined by
\begin{equation}
c^{ab} = \left\{ 
\begin{array}{cc}
1 & \text{if $a = b = +$}, \\
-1 & \text{if $a = b = -$}, \\
0 & \text{otherwise}. \end{array}
\right.
\end{equation}
%In Eq.~(\ref{eq-zgf}), the time branch indices have been suppressed on the 
%metric tensor, for simplicity of notation.  It will be clear how to reinstate
%the time branch indices on the metric tensor in the CTP--2PI effective action,
%order by order in the loop expansion. 
The generating functional for normalized $n$-point functions is
\begin{equation}
W[J,K,g^{\mu\nu}] = -i\hbar \text{ln} Z[J,K,g^{\mu\nu}],
\label{eq-wgf}
\end{equation}
in terms of which we can define the classical scalar field on ${\mathcal M}$
\begin{equation}
\label{eq-mf}
\phih_a(x)_{\text{{\tiny $JK$}}} = c_{ab} \frac{1}{\sqrt{-g}} \frac{
\delta W[J,K,g^{\mu\nu}]}{\delta J_b(x)}
\end{equation}
and the scalar two-point function on ${\mathcal M}$ in the presence of
the sources $J_a$ and $K_{ab}$,
\begin{equation}
\label{eq-gf}
\hbar G_{ab}(x,x')_{\text{{\tiny $JK$}}} =
2 c_{ac} c_{bd} \frac{1}{\sqrt{-g}} \frac{1}{\sqrt{-g'}} \frac{\delta W[J,K
g^{\mu\nu}]}{\delta K_{cd}(x,x')} - \phih_a(x) \phih_b(x'),
\end{equation}
where the $JK$ subscripts indicate that $\phih$ and $G$ are functionals
of the $J_a$ and $K_{ab}$ sources.  In the limit $J_a = K_{ab} = 0$, the 
classical field is the same on the two time branches,
\begin{equation}
\left.\phih_{+}\right|_{J=K=0} = 
\left.\phih_{-}\right|_{J=K=0} = 
\langle \Omega | \PhiH | \Omega \rangle \equiv \phih
\end{equation}
and it is equivalent to the mean field $\phih$.  In the same limit, $G_{ab}$ 
becomes the CTP propagator for the fluctuation field defined in
Eq.~(\ref{eq-ff}),
\begin{mathletters}
\begin{eqnarray}
\label{eq-tpfa}
&&
\left. \hbar G_{++}(x,x')\right|_{J=K=0} = 
\langle \Omega | T(\vphiH(x)\vphiH(x'))|\Omega
\rangle,\\
&&
\left. \hbar G_{--}(x,x')\right|_{J=K=0} = 
\langle \Omega | \tilde{T}(\vphiH(x)\vphiH(x'))
|\Omega\rangle, \\
&&
\left. \hbar G_{+-}(x,x')\right|_{J=K=0} = 
\langle \Omega|\vphiH(x')\vphiH(x)|\Omega
\rangle,
\\
&&
\left. \hbar G_{-+}(x',x)\right|_{J=K=0} = 
\langle \Omega | \vphiH(x)\vphiH(x')|\Omega\rangle,
\label{eq-tpfd}
\end{eqnarray}
\end{mathletters}
where $T$ denotes temporal ordering and $\tilde{T}$ denotes antitemporal
ordering.  In the coincidence limit\footnote{The variance $\langle \varphi(x)^2
\rangle$ is divergent in four spacetime dimensions, and should be regularized
using a covariant procedure \cite{birrell:1982a,bunch:1979a}.} 
$x'=x$, all four components 
(\ref{eq-tpfa})--(\ref{eq-tpfd}) are equivalent to the variance $\langle \vphiH^2 \rangle$ defined in 
Eq.~(\ref{eq-sv}).  Provided we can invert Eqs.~(\ref{eq-mf}) and 
(\ref{eq-gf}) to obtain $J_a$ and $K_{ab}$ in terms of $\phih_a$ and 
$G_{ab}$, the CTP--2PI effective action can 
be defined as the double Legendre transform (in both $J_a$ and $K_{ab}$) of
$W[J,K,g^{\mu\nu}]$,
\begin{eqnarray}
\Gamma[\phih,G,g^{\mu\nu}] && = W[J,K,g^{\mu\nu}] 
- \int_M d^{\;4}x \sqrt{-g} c^{ab} J_a(x) 
\phih_b(x) \nonumber \\
&& - \frac{1}{2} \int_M d^{\;4}x \sqrt{-g} \int_M d^{\;4}x' \sqrt{-g'}
c^{ab} c^{cd} K_{ac}(x,x') \left[ \hbar G_{bd}(x,x') + \phih_b(x) \phih_d(x')
\right].
\label{eq-ea}
\end{eqnarray}
The inverses of Eqs.~(\ref{eq-mf}) and (\ref{eq-gf})
can be obtained by functional differentiation of Eq.~(\ref{eq-ea})
with respect to $\phih_a$,
\begin{equation}
\frac{1}{\sqrt{-g}} 
\frac{\delta \Gamma [\phih,G,g^{\mu\nu}]}{\delta \phih_a(x)} 
= -c^{ab} J_b(x)_{\text{{\tiny $\phih G$}}} 
- \frac{1}{2} c^{ab} c^{cd} \int_M d^{\;4}x' 
\sqrt{-g'} \left[ K_{bd}(x,x')_{\text{{\tiny $\phih G$}}} +
K_{db}(x',x)_{\text{{\tiny $\phih G$}}}\right]\phih_c(x'),
\label{eq-dj} 
\end{equation}
and with respect to $G_{ab}$,
\begin{equation}
\frac{1}{\sqrt{-g}} \frac{1}{\sqrt{-g'}} 
\frac{\delta \Gamma[\phih,G,g^{\mu\nu}]}{\delta G_{ab}(x,x')} 
= -\frac{\hbar}{2}
c^{ac} c^{bd} K_{cd}(x,x')_{\text{{\tiny $\phih G$}}},
\label{eq-dk}
\end{equation}
where the $\phih G$ subscript indicates that $K_{ab}$ and $J_a$ are
functionals of $\phih_a$ and $G_{ab}$.  Inserting Eqs.~(\ref{eq-dj}) and
(\ref{eq-dk}) into Eq.~(\ref{eq-ea}) yields a functional integrodifferential
equation for the CTP--2PI effective action in terms of $\phih$ and $G$ only,
so the $JK$ subscripts can be dropped.  It is useful to change the variable
of functional integration to be the fluctuation field about $\phih_a$,
defined by
\begin{equation}
\varphi_a \equiv \phi_a - \phih_a.
\label{eq-fcv}
\end{equation}
Performing the change-of-variables $D\phi \rightarrow D\varphi$, the
equation for $\Gamma$ is
\begin{eqnarray}
\Gamma[\phih,G,g^{\mu\nu}] 
&& = \int_M d^{\;4}x \int_M d^{\;4}x' \frac{\delta \Gamma
[\phih,G]}{\delta G_{ba}(x',x)}G_{ab}(x,x')  \nonumber \\
&& - i\hbar \text{ln} \Biggl\{ \int_{\text{{\tiny ctp}}} D\varphi_{+}
D\psib_{+} D\psi_{+} D\varphi_{-} D\psib_{-} D\psi_{-} \exp
\biggl[ \frac{i}{\hbar} \Bigl( {\mathcal S}^{\text{{\tiny F}}}[
\varphi + \phih, \psib, \psi, g^{\mu\nu}] \nonumber \\
&& - \int_M d^{\;4}x \frac{\delta \Gamma[\phih,G,g^{\mu\nu}]}{\delta \phih_a} 
\varphi_a
- \frac{1}{\hbar} \int_M d^{\;4}x \int_M d^{\;4}x' \frac{\delta \Gamma
[\phih,G,g^{\mu\nu}]}{\delta G_{ba}(x',x)} \varphi_a(x) \varphi_b(x') \Bigr) 
\biggr]\Biggr\},
\end{eqnarray}
which has the formal solution
\begin{eqnarray}
\Gamma[\phih,G,g^{\mu\nu}] = 
&& {\mathcal S}^{\phi}[\phih] - \frac{i\hbar}{2} \text{ln}\, \text{det} G_{ab} 
- i\hbar \text{ln} \, \text{det} F_{ab} 
+ \Gamma_2[\phih,G] \nonumber \\
&& + \frac{i\hbar}{2} \int_M d^{\;4}x \sqrt{-g} \int_M d^{\;4}x'
\sqrt{-g'} {\mathcal A}^{ab} (x',x) G_{ab}(x,x'),
\label{eq-fs}
\end{eqnarray}
where ${\mathcal A}^{ab}$ is the second functional derivative of the 
scalar part of the classical action $S^{\phi}$, evaluated at $\phih$,
\begin{eqnarray}
i {\mathcal A}^{ab}(x,x') &=& \frac{1}{\sqrt{-g}}
\left( \frac{\delta^2 {\mathcal S}^{\phi}}{\delta \phi_a(x) \delta \phi_b(x')}
[\phih] \right) \frac{1}{\sqrt{-g'}} \nonumber \\ & = & 
-\left[ 
c^{ab} ( \square_x + m^2 + \xi R(x) )
+ c^{abcd} \frac{\lambda}{2} \phih_c(x) \phih_d(x) 
\right] \delta(x-x') \frac{1}{\sqrt{-g'}},
\label{eq-iolscp}
\end{eqnarray}
The symbol $F_{ab}$ denotes the one-loop CTP spinor propagator, 
which is defined by
\begin{equation}
F_{ab}(x,x') \equiv {\mathcal B}_{ab}^{-1}(x,x'),
\label{eq-olspp}
\end{equation}
where we are suppressing spinor indices, and the inverse spinor propagator
${\mathcal B}^{ab}$ is defined by
\begin{eqnarray}
i {\mathcal B}^{ab}(x,x') &=& \frac{1}{\sqrt{-g}} \left[
\frac{\delta^2 ({\mathcal S}^{\psi}[\psib,\psi] + 
{\mathcal S}^{\text{{\tiny Y}}}[\psib,\psi;\phih])}{\delta \psi_a(x) 
\delta \psib_b(x')} \right] \frac{1}{\sqrt{-g'}} \\
&=&
\left( c^{ab}(i \gamma^{\mu} \nabla'_{\mu} - \mu) - c^{abc} \yc \phih_c(x')
\right) \delta(x'-x)
\frac{1}{\sqrt{-g}} {1}_{\text{{\tiny sp}}}.
\label{eq-iolspp}
\end{eqnarray}
It is clear from Eq.~(\ref{eq-iolspp}) that the use of the
one-loop spinor propagators in the construction of the CTP-2PI-CGEA 
represents a nonperturbative resummation in the Yukawa 
coupling constant, which (as discussed above) goes beyond the standard 
time-dependent perturbation theory.
The boundary conditions which define the inverses of Eqs.~(\ref{eq-iolscp})
and (\ref{eq-iolspp}) are the boundary conditions at the initial data surface
in the functional integral in Eq.~(\ref{eq-zgf}), which in turn, define the 
initial quantum state $|\Omega\rangle$.  The one-loop spinor propagator 
$F_{ab}$ is related to the expectation values of the spinor Heisenberg 
field operators for a spinor field in the presence of the c-number background
field $\hat{\phi}$,
\begin{mathletters}
\begin{eqnarray}
&&\hbar \left.F_{++}(x,x')\right|_{\phihp = \phihm = \phih}
 = \langle \Omega | T(\PsiH(x) \PsibH(x'))|\Omega \rangle, \\
&&\hbar \left.F_{--}(x,x')\right|_{\phihp = \phihm = \phih}
 = \langle \Omega | \tilde{T} (\PsiH(x) \PsibH(x'))|\Omega
\rangle, \\
&&\hbar \left.F_{+-}(x,x')\right|_{\phihp = \phihm = \phih}
 = -\langle \Omega | \PsibH(x') \PsiH(x) | \Omega \rangle, \\
&&\hbar \left.F_{-+}(x,x')\right|_{\phihp = \phihm = \phih}
 = \langle \Omega | \PsiH(x) \PsibH(x') | \Omega \rangle,
\end{eqnarray}
\end{mathletters}
where the spinor Heisenberg field operators obey the equations
\begin{mathletters}
\begin{eqnarray}
&& (i \gamma^{\mu} \nabla_{\mu} - \mu - f \phih ) \Psi = 0, \\
&& (-i \gamma^{\mu} \nabla_{\mu} - \mu - f \phih ) \bar{\Psi} = 0.
\end{eqnarray}
\end{mathletters}
The CTP spinor propagator components satisfy the relations 
(valid only when $\phihp = \phihm = \phih$)
\begin{mathletters}
\begin{eqnarray}
&& F_{++}(x,x')^{\dagger} = F_{--}(x',x),\\
&& F_{--}(x,x')^{\dagger} = F_{++}(x',x),\\
&& F_{-+}(x,x')^{\dagger} = F_{-+}(x',x),\\
&& F_{+-}(x,x')^{\dagger} = F_{+-}(x',x),
\end{eqnarray}
\end{mathletters}
The functional $\Gamma_2[\phih,G]$ is defined as $-i\hbar$ times the sum of 
all vacuum diagrams drawn according to the following rules:
\begin{enumerate}
\item{Vertices carry spacetime ($x \in M$) and time branch ($a \in \{+,-\}$)
labels.}
\item{Scalar field lines denote $\hbar G_{ab}(x,x')$.}
\item{Spinor lines denote the one-loop CTP
spinor propagator $\hbar F_{ab}(x,x')$ (spinor indices are suppressed), defined 
in Eq.~(\ref{eq-olspp}).}
\item{There are three interaction vertices, given by $i {\mathcal S}^{
\text{{\tiny I}}} /\hbar$, which is defined by
\begin{mathletters}
\begin{eqnarray} 
&& {\mathcal S}^{\text{{\tiny I}}}[\phih,\varphi,\psib,\psi] =
S^{\text{{\tiny I}}}[\phihp,\vphip,\psibp,\psip] -
S^{\text{{\tiny I}}}[\phihm,\vphim,\psibm,\psim], \\
&& S^{\text{{\tiny I}}}
[\phih,\vphi,\psib,\psi] = -\int d^4 x \sqrt{-g} \left[
f \vphi \psib \psi + \frac{\lambda}{24} \vphi^4 +
\frac{\lambda}{6} \phih \vphi^3\right],
\end{eqnarray}
\end{mathletters}
where we have followed the notation of Eq.~(\ref{eq-syctp}).}
\item{Only diagrams which are two-particle-irreducible 
with respect to cuts of {\em scalar\/} lines contribute to $\Gamma_2$.}
\end{enumerate}
The distinction between the CTP-2PI, {\em coarse-grained\/}
effective action which arises here, and the fully
two-particle-irreducible effective action (2PI with respect to scalar 
{\em and\/} spinor cuts), is due to the fact that we only Legendre-transformed
sources coupled to $\phi$; i.e., the spinor field is treated as the 
environment.  In Eq.~(\ref{eq-iolscp}), the curved-spacetime
Dirac $\delta$ function is defined as in \cite{birrell:1982a}. Comparison of 
Eq.~(\ref{eq-fs}) above with Eq.~(4.13) of Ref.~\cite{ramsey:1997a}
[which was computed for the $O(N)$
model] shows that the $\text{tr}\,\text{ln}F_{ab}$ 
in Eq.~(\ref{eq-fs}) differs from the usual one-loop term by a factor of 2,
owing to the difference (in the exponent) between the Gaussian integral 
formulas for real and complex fields \cite{cornwall:1974a}.  

\begin{figure}[htb]
\begin{center}
\epsfig{file=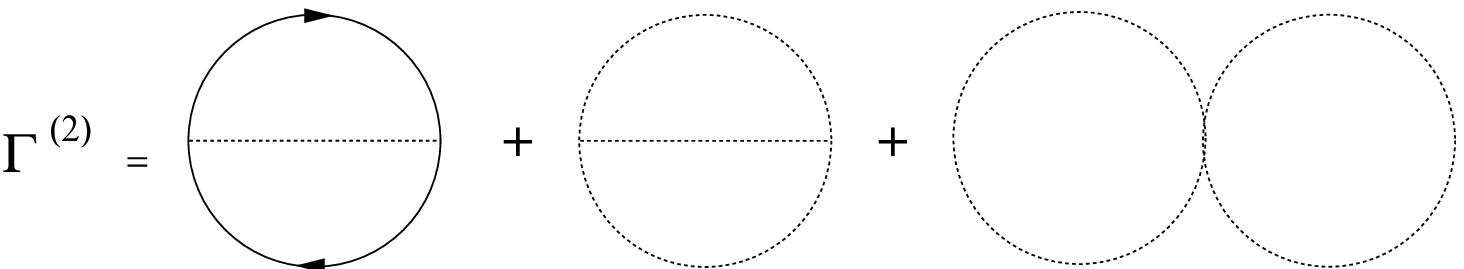,width=4.0in}
\end{center}
\caption{Diagrammatic expansion for $\Gamma^{(2)}$, the two-loop
part of the CTP-2PI-CGEA.  Solid lines represent the spinor propagator $F$
(as defined in Sec.~\ref{sec-necgidcst}), and dotted lines represent the
scalar propagator $G$.  The vertices terminating three $\phi$ lines
are proportional to the scalar mean field $\phih$.  Each vertex carries
spacetime $(x)$ and CTP $(+,-)$ labels.}
\label{fig-gam2}
\end{figure}

The functional $\Gamma_2[\phih,G,g^{\mu\nu}]$ can be evaluated in a loop 
expansion, which corresponds to an expansion in powers of $\hbar$,
\begin{equation}
\Gamma_2[\phih,G,g^{\mu\nu}] = \sum_{l = 2}^{\infty} \hbar^l 
\Gamma^{(l)}[\phih,G,g^{\mu\nu}],
\end{equation}  
starting with the two-loop term, $\Gamma^{(2)}$, which has a diagrammatic
expansion shown in Fig.~\ref{fig-gam2}.
The $\lambda \phi^4$ self-interaction leads to two terms in the two-loop
part of the effective action, the second and third graphs
of Fig.~\ref{fig-gam2}.  They are the ``setting sun'' diagram,
which is $O(\lambda^2)$, and the ``double bubble,'' which is
$O(\lambda)$, respectively.   
The Yukawa interaction leads to only one diagram in 
$\Gamma^{(2)}$, the first diagram of Fig.~\ref{fig-gam2},
\begin{equation}
\frac{i f^2}{2} c^{aa'a''} c^{bb'b''} 
\int d^{\,4}x \sqrt{-g} \int d^{\,4}x' \sqrt{-g'} G_{ab}(x,x') 
\text{tr}_{\text{{\tiny sp}}}\left[ F_{a'b'}(x,x') F_{b''a''}(x',x)\right],
\label{eq-dd2}
\end{equation}
where the trace is understood to be over the spinor indices which are not
shown, and the three-index symbol $c^{abc}$ is defined by
\begin{equation}
c^{abc} = \left\{ 
\begin{array}{cc}
1 & \text{if $a = b = c = +$,} \\
-1 & \text{if $a = b = c = -$,} \\
0 & \text{otherwise.}
\end{array}
\right.
\end{equation}
Here, we treat the $\lambda$ self-interaction using 
 the time-dependent Hartree-Fock approximation 
\cite{cornwall:1974a}, which is equivalent to retaining the
$O(\lambda)$ (double-bubble) graph and dropping the $O(\lambda^2)$
(setting sun) graph.  We assume for the present study 
that the coupling $\lambda$ is sufficiently
small that the $O(\lambda^2)$ diagram is unimportant on the
time scales of interest in the fermion production regime of the inflaton
dynamics.  The mean-field and gap equations including both the
$O(\lambda)$ and the $O(\lambda^2)$ diagrams have been derived in a
general curved spacetime in our previous paper \cite{ramsey:1997a}.

\subsection{Evolution equations for $\phih$ and $G$ in curved spacetime}
The (bare) semiclassical field equations for the two-point function, 
mean field, and metric can be obtained from the CTP-2PI-CGEA by 
functional differentiation with respect $G_{ab}$, $\phih_a$, and $g^{\mu\nu},$
followed by identifications of $\phih$ and $g^{\mu\nu}$ on the two time
branches \cite{ramsey:1997b},
\begin{mathletters}
\begin{eqnarray}
\label{eq-seea}
\left.
\frac{\delta {\mathcal S}^{\text{{\tiny G}}}[g^{\mu\nu}] + \Gamma[\phih,G,
g^{\mu\nu}]}{\delta g^{\mu\nu}_a}\right|_{
\phih_{+} = \phih_{-} \equiv \phih; \;\;\;\;
g^{\mu\nu}_{+} = g^{\mu\nu}_{-} \equiv g^{\mu\nu}} & = & 0, \\
\left.\frac{\delta \Gamma[\phih,G,g^{\mu\nu}]}{\delta \phih_a}\right|_{
\phih_{+} = \phih_{-} \equiv \phih; \;\;\;\;
g^{\mu\nu}_{+} = g^{\mu\nu}_{-} \equiv g^{\mu\nu}} & = & 0, 
\label{eq-seeb} \\
\left.\frac{\delta \Gamma[\phih,G,g^{\mu\nu}]}{\delta G_{ab}}\right|_{
\phih_{+} = \phih_{-} \equiv \phih; \;\;\;\;
g^{\mu\nu}_{+} = g^{\mu\nu}_{-} \equiv g^{\mu\nu}} & = & 0.
\label{eq-seec}
\end{eqnarray}
\end{mathletters}
Equations~(\ref{eq-seea})--(\ref{eq-seec}) constitute the semiclassical 
approximation to the full quantum dynamics for the system described by
the classical action (\ref{eq-ca}).  Equation~(\ref{eq-seea}) should be 
understood as following after time branch indices have been reinstated on
the metric tensor in the CTP-2PI-CGEA \cite{ramsey:1997a}.
The field equation of semiclassical gravity (with bare parameters) is obtained
directly from Eq.~(\ref{eq-seec}),
\begin{equation}
G_{\mu\nu} + \Lambda_c g_{\mu\nu} + c \, ^{(1)}H_{\mu\nu} + b \,
^{(2)}H_{\mu\nu} = -8\pi G \langle T_{\mu\nu} \rangle,
\label{eq-see}
\end{equation}
in terms of the (unrenormalized) quantum energy-momentum tensor defined by
\begin{equation}
\langle T_{\mu\nu} \rangle = \frac{2}{\sqrt{-g}} \left. \frac{\delta
\Gamma[\phih,G,g^{\mu\nu}]}{\delta g^{\mu\nu}_{+}}\right|_{
\phih_{+} = \phih_{-} = \phih \;\;\;\; g^{\mu\nu}_{+} =
g^{\mu\nu}_{-} = g^{\mu\nu}}.
\end{equation}
The energy-momentum tensor $\langle T_{\mu\nu}\rangle$ is divergent in
four spacetime dimensions, and must be regularized via a covariant procedure
\cite{birrell:1982a,ramsey:1997b}.

Making the two-loop approximation to the CTP-2PI-CGEA,
where we take $\Gamma_2 \simeq \hbar^2 \Gamma^{(2)}$, and dropping the
$O(\lambda^2)$ diagram from $\Gamma_2$, the mean-field equation 
(\ref{eq-seeb}) becomes
\begin{equation}
\left( \square + m^2 + \xi R(x) + \frac{\lambda}{6} \phih^2(x) +
\frac{\lambda\hbar}{2} G(x,x) \right) 
\phih + \hbar f \text{tr}_{\text{{\tiny sp}}} [F_{ab}(x,x)] - \hbar^2 g^3 
\Sigma(x) = 0,
\label{eq-npmfe}
\end{equation}
where $G(x,x)$ is the coincidence limit of $G_{ab}(x,x')$, and
in terms of a function $\Sigma(y)$ defined by
\begin{eqnarray}
\Sigma(y) \equiv \int d^{\,4}x \sqrt{-g} \int d^{\,4}x' \sqrt{-g'} 
\biggl\{ && 
G_{++}(x,x') \text{tr}_{\text{{\tiny sp}}}\left[F_{++}(x,y)
                                           F_{++}(y,x') F_{++}(x',x)\right] 
\nonumber \\
- &&
G_{-+}(x,x') \text{tr}_{\text{{\tiny sp}}}\left[F_{-+}(x,y) 
                                           F_{++}(y,x') F_{+-}(x',x)\right] 
\nonumber \\
- &&
G_{+-}(x,x') \text{tr}_{\text{{\tiny sp}}}\left[F_{++}(x,y) 
                                           F_{+-}(y,x') F_{-+}(x',x)\right] 
\nonumber \\
+ &&
G_{--}(x,x') \text{tr}_{\text{{\tiny sp}}}\left[F_{-+}(x,y)
                                           F_{+-}(y,x') F_{--}(x',x)\right] 
\biggr\}.
\label{eq-sigma}
\end{eqnarray}
Making use of the curved spacetime definitions of the scalar and 
spinor field Hadamard kernels \cite{birrell:1982a}
\begin{mathletters}
\begin{eqnarray}
G^{(1)}(x,x') &=& \langle \Omega | \{ \varphi_{\text{{\tiny H}}}(x), 
\varphi_{\text{{\tiny H}}}(x') \} | \Omega \rangle, \\
F^{(1)}(x,x') &=& \langle \Omega | [ \Psi_{\text{{\tiny H}}}(x), 
\bar{\Psi}_{\text{{\tiny H}}}(x')] | \Omega \rangle,
\label{eq-dshf}
\end{eqnarray}
\end{mathletters}
and retarded propagators
\begin{mathletters}
\begin{eqnarray}
G_R(x,x') &=& i \theta(x,x') \langle \Omega | [ \varphi_{\text{{\tiny H}}}(x),
\varphi_{\text{{\tiny H}}}(x') ] | \Omega \rangle, \\
F_R(x,x') &=& i \theta(x,x') \langle \Omega | \{ \Psi_{\text{{\tiny H}}}(x),
\bar{\Psi}_{\text{{\tiny H}}}(x') \} | \Omega \rangle, 
\label{eq-dsrp}
\end{eqnarray}
\end{mathletters}
the function $\Sigma(y)$ can be recast in a manifestly real and causal form,
\begin{eqnarray}
\Sigma(y) = -2 \int d^4 x \sqrt{-g} \int d^4 x' \sqrt{-g'} \, \text{Re} \,
\text{tr}_{\text{{\tiny sp}}} && \Bigl[\Bigl( \theta(x,x') G^{(1)}(x',x) 
F^{(1)}(x,x') \nonumber \\ && -
G_R(x,x')^{\star} F_R(x,x') \Bigr) F_R(y,x')^{\star} F_R(y,x) \Bigr],
\end{eqnarray}
from which it is clear that the integrand vanishes whenever $x$ or $x'$ is
to the future of $y$.
The ``gap'' equation for $G_{ab}$ is obtained from Eq.~(\ref{eq-seec}), 
\begin{equation}
(G^{-1})^{ba}(x,x') = {\mathcal A}^{ba}(x,x') + 
\frac{i\lambda\hbar}{2} c^{ba} G(x,x)
\delta(x-x')\frac{1}{\sqrt{-g'}}
+ \hbar \yc^2 c^{aa'a''}c^{bb'b''}
\text{tr}_{\text{{\tiny sp}}} \left[ F_{a'b'}(x,x') F_{b''a''}(x',x) \right].
\label{eq-fge}
\end{equation}
Multiplying Eq.~(\ref{eq-fge}) through by $G_{ab}$, performing a spacetime
integration, and taking the $++$ component, we obtain
\begin{eqnarray}
\biggl(( \square + m^2 + \xi R + \frac{\lambda}{2} \phih^2 + && 
\frac{\lambda \hbar}{2} G(x,x) \biggr) G_{++}(x,x')  \nonumber \\
&& + \hbar \yc^2 \int dx'' 
\sqrt{-g''} {\mathcal K}(x,x'') G_{++}(x'',x') = -i \delta (x-x')
\frac{1}{\sqrt{-g'}},
\label{eq-dge}
\end{eqnarray}
in terms of a kernel ${\mathcal K}(x,x'')$ defined by
\begin{equation}
{\mathcal K}(x,x') = -i\text{tr}_{\text{{\tiny sp}}}
\left[ F_{++}(x,x') F_{++}(x',x) - F_{+-}(x,x')F_{-+}(x',x) 
\right].
\end{equation}
Making use of Eqs.~(\ref{eq-dshf}) and (\ref{eq-dsrp}), this kernel takes the
form
\begin{equation}
{\mathcal K}(x,x') = \text{Re} \, \text{tr}_{\text{{\tiny sp}}} \left[ F_R(x,x') 
F^{(1)}(x',x) \right],
\label{eq-kapker}
\end{equation}
which shows that the gap equation (\ref{eq-dge}) is manifestly real and causal.
As will be shown below in Sec.~\ref{sec-dsasfo} (in a perturbative limit), 
the kernel ${\mathcal K}(x,x')$ is
dissipative, and it reflects back reaction from fermionic particle production
induced by the time-dependence of the inflaton {\em variance}\@.  
The gap equation (\ref{eq-dge}) is therefore damped for modes above 
threshold\footnote{See \cite{calzetta:1989b} for a similar discussion in the 
context of spinodal decomposition in quantum field theory}, and this damping 
is not accounted for in the 
1PI treatments of inflaton dynamics (where only the inflaton mean field is 
dynamical).  In contrast to previous studies 
\cite{boyanovsky:1995d,boyanovsky:1995b,boyanovsky:1995a} 
which assumed a local equation of motion for the inflaton
propagator, the two-loop gap equation obtained from the CTP-2PI-CGEA includes
a {\em nonlocal\/} kernel, which is a generic feature of back reaction from 
particle
production.  As stressed above, the dissipative dynamics of the inflaton
two-point function can be important when the inflaton variance is on the
order of the square of the inflaton mean-field amplitude; such conditions 
may exist at the end of preheating.

The set of evolution equations (\ref{eq-npmfe}) for $\phih$ and (\ref{eq-dge})
for $G$, is formally complete to two loops.  Dissipation arises due to the
coarse-graining of the spinor degrees of freedom.  These dynamical 
equations are useful for general purposes, and are valid in a general 
background spacetime.  However, in order to get explicit results, one needs to
introduce approximations, as we now do.  

\section{Dynamics of small-amplitude inflaton oscillations}
\label{sec-dsasfo}
The effective evolution equations for the inflaton mean field $\phih$
and variance $\langle \varphi^2 \rangle$
derived in the previous section are useful for studying fermion production 
when $\phih_0$, the amplitude of the spatially homogeneous
inflaton mean-field oscillations, is large, and the inflaton variance
is of the same order-of-magnitude as $(\phih_0)^2$.
 As discussed in Sec.~\ref{sec-necgidcst} above, such conditions can 
prevail at the end of the preheating period in chaotic
inflation with unbroken symmetry \cite{boyanovsky:1996b,ramsey:1997b}.
Because of the dissipative kernel ${\mathcal K}(x,x')$ in the gap equation
(\ref{eq-dge}), which damps
the evolution of $G$, and the back reaction terms in the mean-field equation,
which damp the oscillations of $\phih$, eventually the condition,
\begin{equation}
f \phih_0 \ll m
\label{eq-vpt}
\end{equation}
will hold,
%(where $\phih_0$ is the slowly-varying
%amplitude of oscillation of the spatially homogeneous inflaton
%mean field), 
at which point it is justifiable to follow the mean-field dynamics
using the perturbative, 1PI, coarse-grained effective action 
\cite{shtanov:1995a}.  Although in principle
one should study this process in a general curved spacetime, 
for simplicity we assume spatial homogeneity, and that the inflaton mass is 
much greater than the Hubble constant, $m \gg H$.  While this condition alone
is in general {\em not\/} sufficient to ensure that curved spacetime 
effects are negligible during reheating
(see, for example, \cite{ramsey:1997b}, where cosmic expansion {\em does\/}
affect preheating dynamics even though $m \gg H$), with the additional 
assumption of condition (\ref{eq-vpt}) it is reasonable to neglect the 
effect of cosmic expansion in the {\em spinor\/} propagators 
\cite{boyanovsky:1995c,boyanovsky:1995d}.  In this and the following section,
we also neglect the self-coupling $\lambda$, because for the case of unbroken
symmetry, the lowest-order $\lambda$-dependent contribution to the 
perturbative inflaton self-energy is $O(\hbar^2)$ \cite{boyanovsky:1995a}, 
and we are only concerned with one-loop dynamics in this section.  

\begin{figure}[htb]
\begin{center}
\epsfig{file=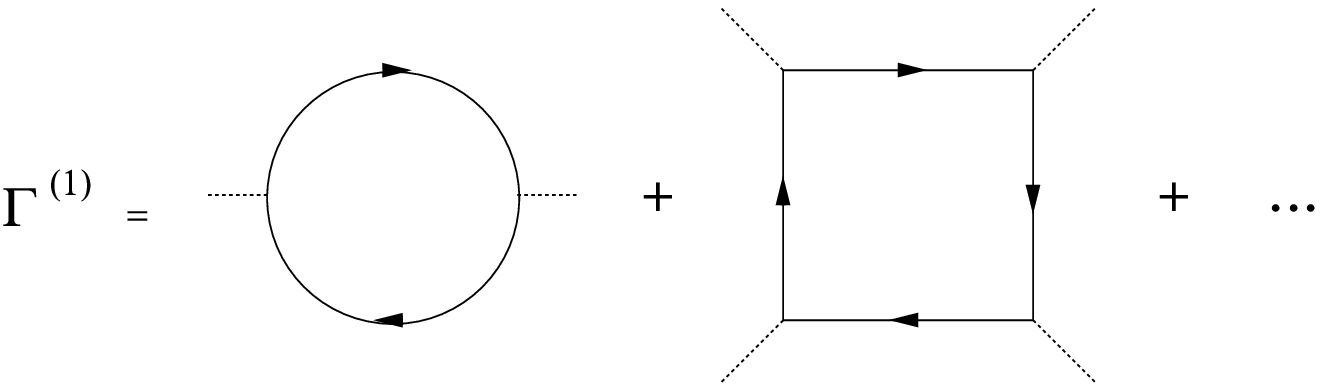,width=3.5in}
\end{center}
\caption{Diagrammatic expansion for $\Gamma^{(1)}$, the one-loop
part of the CTP-1PI-CGEA.  Solid lines represent the spinor propagator 
$F$ (as defined in Sec.~\ref{sec-dsasfo}), and dotted lines represent 
multiplication by the scalar mean field $\phih$.  Each vertex carries
spacetime $(x)$ and CTP $(+,-)$ labels.}
\label{fig-gam1}
\end{figure}

Let us therefore specialize to Minkowski space, and implement a perturbative
expansion of the CTP effective action in powers of the mean field $\phih$.
This formally entails a solution of the gap equation (\ref{eq-dge}) 
for $G$, a back-substitution of the solution into the CTP-2PI 
coarse-grained effective action, and a subsequent expansion of this
expression (now a functional of $\phih$ only) in powers of $\phih$.
The resulting perturbative expansion for the effective action 
contains only free-field  propagators.  For consistency, one should
use an initial density matrix for the spinor degrees of freedom which
corresponds to the end-state particle occupation numbers of the 
nonperturbative dynamics of Sec.~\ref{sec-necgidcst}.  For simplicity,
however, we assume the initial quantum state for the spinor field is the
vacuum state.  Hereafter, $F_{ab}$ denotes the free-field,
Minkowski-space, vacuum spinor CTP propagator, whose components are given by
\cite{chou:1985a,stylianopoulos:1991a,cooper:1994a}
\begin{mathletters}
\begin{eqnarray}
F_{++}(x,x') &=& \int \frac{d^{\,4}p}{(2\pi)^4} e^{-i p (x-x')} \frac{i (\not{p}
+ m)}{p^2 - \mu^2 + i\epsilon}, \\
F_{--}(x,x') &=& -\int \frac{d^{\,4}p}{(2\pi)^4} e^{-i p (x-x')} \frac{i (\not{p} 
+ m)}{p^2 - \mu^2 - i\epsilon}, \\
F_{-+}(x,x') &=& \int \frac{d^{\,4}p}{(2\pi)^4} e^{-i p (x-x')} 2\pi (\not{p} 
+ m) \delta (p^2 - \mu^2) \theta(p^0), \\
F_{+-}(x,x') &=& \int \frac{d^{\,4}p}{(2\pi)^4} e^{-i p(x-x')} 2\pi (\not{p} 
+ m) \delta (p^2 - \mu^2) \theta(-p^0). 
\end{eqnarray}
\end{mathletters}
The $++$ and $--$ propagators admit a representation in terms of
a time-ordering function $\theta(x,y) = \theta(x^0 - y^0)$,
\begin{mathletters}
\begin{eqnarray}
F_{++}(x,x') &&= \theta(x,x') F_{-+}(x,x') + \theta(x',x) F_{+-}(x,x'),\\
F_{--}(x,x') &&= \theta(x,x') F_{+-}(x,x') + \theta(x',x) F_{-+}(x,x').
\end{eqnarray}
\end{mathletters}
The CTP effective action can be expanded in powers of $f^2$, and we find
\begin{equation}
\Gamma[\phih] = {\mathcal S}^{\phi}[\phih] 
- \frac{i\hbar}{2} \text{ln}\,\text{det} (\tilde{{\mathcal A}}^{ab})^{-1}
- i\hbar \text{ln}\,\text{det} F_{ab}
+ \Gamma_1[\phih],
\end{equation}
where the kernel $\tilde{{\mathcal A}}$ is defined by
\begin{equation}
i \tilde{{\mathcal A}}^{ab}(x,x') = -c^{ab}(\square_x + m^2)\delta(x-x'),
\end{equation}
%where the kernel ${\mathcal A}^{ab}$ was defined in Eq.~(\ref{eq-iolscp})
%(noting that $R = 0$ in Minkowski space), 
and $\Gamma_1$ is defined as
$-i\hbar$ times the sum of all one-particle-irreducible diagrams constructed
with lines given by $\hbar \tilde{{\mathcal A}}^{-1}$ and 
$\hbar F_{ab}$, and vertices given
by ${\mathcal S}^{\text{{\tiny Y}}}[\phih,\psib,\psi]/\hbar$ and
${\mathcal S}^{\text{{\tiny Y}}}[\varphi,\psib,\psi]/\hbar$.
Because the free-field propagators $\tilde{{\mathcal A}}^{-1}$ and 
$F_{ab}$ do not depend on $\phih$, the $\log(\det)$ 
do not contribute to the variation of $\Gamma[\phih]$
with respect to $\phih$, and therefore, they can be dropped.
The functional $\Gamma_1[\phih]$ can be expanded in powers of $\hbar$,
\begin{equation}
\Gamma_1[\phih] = \sum_{l=1}^{\infty} \hbar^l \Gamma^{(l)}[\phih],
\end{equation}
where the term $\Gamma^{(l)}[\phih]$ is the sum of all 1PI $l$-loop graphs.
Order by order in the loop expansion and the coupling constant, 
the CTP 1PI effective action must satisfy the unitarity condition 
\begin{equation}
\left.\Gamma_1\right|_{\phihp = \phih; \;\; \phihm = \phih} = 0,
\label{eq-uc}
\end{equation}
which has been verified to two-loop order in the case of 
scalar $\lambda \phi^4$ field theory \cite{jordan:1986a}.
The one-loop term in the loop expansion of the CTP effective action,
$\Gamma^{(1)}[\phih]$, can be further expanded in powers of $f^2$,
\begin{equation}
\Gamma^{(1)}[\phih] = \sum_{n=1}^{\infty} f^{2n} \Gamma^{(1)}_{2n}[\phih],
\end{equation}
which corresponds to the usual amplitude expansion of the CTP
effective action \cite{boyanovsky:1995d}.  Figure~\ref{fig-gam1} shows the
diagrammatic expansion of $\Gamma^{(1)}$. The terms $\Gamma_{2n}^{(1)}[\phih]$
are generally divergent, but since the theory is renormalizable in the 
standard ``in-out'' formulation, it is renormalizable in the closed-time-path,
``in-in'' formulation \cite{jordan:1986a,calzetta:1987a}.

\subsection{One-loop perturbative effective action at $O(f^2)$}
\label{sec-olpefa}
The $O(f^2)$ term in the expansion of the one-loop CTP effective action,
which is the first term in Fig.~\ref{fig-gam1},
takes the form\footnote{Note that there are no nonzero graphs with 
an odd number of vertices in this model.}  
\begin{equation}
\Gamma^{(1)}_2[\phih] = -\frac{i}{2}
c^{abc} c^{a'b'c'} \int d^{\,4}x d^{\,4}x' \phih_a(x)
\phih_{a'}(x') F_{bb'}(x,x') F_{c'c}(x',x).
\end{equation}
Making use of sum and difference variables
\begin{mathletters}
\begin{eqnarray}
&& \Sigma(x) = \frac{1}{2} \left[ \phih_{+}(x) + \phih_{-}(x') \right], 
\label{eq-sdva} \\
&& \Delta(x) = \phih_{+}(x) - \phih_{-}(x'),
\label{eq-sdvb}
\end{eqnarray}
\end{mathletters}
the functional $\Gamma^{(1)}_2[\phih]$ can be recast in the form
\begin{equation}
\Gamma^{(1)}_2[\phih] = \int d^{\,4}x d^{\,4}x' \biggl[ \Sigma(x) \Delta(x')
{\mathcal D}_2(x,x') + \frac{i}{2} \Delta(x) \Delta(x')
{\mathcal N}_2(x,x') \biggr],
\label{eq-d1}
\end{equation}
in terms of manifestly real kernels ${\mathcal D}_2(x,x')$ and 
${\mathcal N}_2(x,x')$ defined by
\begin{mathletters}
\begin{eqnarray}
&&{\mathcal D}_2(x,x') = \,\text{Im} \, \text{tr}_{\text{{\tiny sp}}}
[F_{++}(x,x')F_{++}(x',x) +
F_{+-}(x,x')F_{-+}(x',x)], \label{eq-dk2} \\
&&{\mathcal N}_2(x,x') = -\,\text{Re}\,
\text{tr}_{\text{{\tiny sp}}}[F_{++}(x,x')F_{++}(x',x)].
\label{eq-dn2}
\end{eqnarray}
\end{mathletters}
Only the kernel ${\mathcal D}_2(x,x')$ contributes to the mean-field equation 
of motion.  The kernel ${\mathcal N}_2(x,x')$ constitutes a 
correlator for noise, and will be discussed in Sec.~\ref{sec-nk}.
The unitarity condition (\ref{eq-uc}) requires that the sum of
diagrams proportional to $\Sigma(x)\Sigma(x')$ vanish identically. 
With the definitions of the retarded spinor propagator, Eq.~(\ref{eq-dshf}),
and the spinor Hadamard kernel, Eq.~(\ref{eq-dsrp}),
which in Minkowski space take the form
\begin{mathletters}
\begin{eqnarray}
&& F_R(x,x') = i \theta(x,x') [ F_{-+}(x,x') - F_{+-}(x,x') ], \\
&& F^{(1)}(x,x') = F_{-+}(x,x') + F_{+-}(x,x'),
\end{eqnarray}
\end{mathletters}
the kernel ${\mathcal D}_2(x,x')$ can be written in a manifestly causal
form,
\begin{equation}
{\mathcal D}_2(x,x') = \frac{1}{2}
\, \text{Re} \, \text{tr}_{\text{{\tiny sp}}} [F_R(x,x') F^{(1)}(x',x)].
\label{eq-d2rp}
\end{equation}
Using Eq.~(\ref{eq-d2rp}), 
one can compute ${\mathcal D}_2(x,x')$ in an arbitrary curved
background spacetime.  It should be noted that ${\mathcal D}_2(x,x')$ is
just the lowest-order term in the series expansion of ${\mathcal K}(x,x')$ 
[defined in Eq.~(\ref{eq-kapker})] in 
powers of the coupling constant $f$.  The appearance of the retarded 
propagator in Eq.~(\ref{eq-d2rp}) guarantees that the contribution of
$\Gamma^{(1)}_2$ to the mean-field equation of motion is causal.  

Let us now evaluate ${\mathcal D}_2(x,x')$
using dimensional regularization and the modified minimal subtraction 
($\overline{\text{MS}}$) renormalization prescription \cite{peskin:1995a}.  
Dimensional regularization
requires changing the coupling constant so that the interaction
$S^{\text{{\tiny Y}}}$  has the correct dimensions in $n$ spacetime dimensions,
\begin{equation}
f \rightarrow f \Lambda^{(4-n)/2},
\end{equation}
where we have introduced a parameter $\Lambda$, the renormalization
scale, which has dimensions of mass.   By Lorentz invariance and causality, 
the product of Feynman propagators can be written in terms of an amplitude 
$A_2$ \cite{peskin:1995a},
\begin{equation}
\text{tr}_{\text{{\tiny sp}}} \, [ F_{++}(x,x') F_{++}(x',x)]
= i \int \frac{d^{\,4}k}{(2\pi)^4} e^{-ik(x-x')} A_2(k^2 + i\epsilon),
\label{eq-spp}
\end{equation}
and with this choice of renormalization prescription, the amplitude
$A_2(k^2)$ takes the form
\begin{equation}
A_2(k^2) = -\frac{3}{4\pi^2} \int_0^1 d\alpha E(\alpha;k^2) \text{log}
\, \Biggl( \frac{E(\alpha;k^2)}{\Lambda^2} \Biggr), \label{eq-dm}
\end{equation}
where $E(\alpha;k^2)$ is defined by\footnote{The notation $E(\alpha;k^2)$ 
used here should not be confused with $E(k)$, the complete elliptic integral of
second kind.}
\begin{equation}
E(\alpha;k^2) = \mu^2 - \alpha(1-\alpha)k^2.
\end{equation}
Note that in Eq.~(\ref{eq-dm}), the $\alpha$ integration shows up via the 
Feynman identity
\begin{equation}
\frac{1}{C_1 \ldots C_N} = (N-1)! \int_0^1 d\alpha_1 \cdots
\int_0^1 d\alpha_N \delta(\alpha_1 + \cdots + \alpha_{N - 1}) 
\bigl[ \alpha_1 C_1 + \cdots + \alpha_N C_N \bigr]^{-N}.
\label{eq-fi}
\end{equation}
The $i\epsilon$ appearing in Eq.~(\ref{eq-spp}) ensures that the amplitude 
$A_2$ is evaluated on the physical sheet \cite{eden:1966a,peskin:1995a}.
The logarithm in Eq.~(\ref{eq-dm}) has a negative real argument when
the two conditions $k^2 > 4 \mu^2$ and $|2\alpha-1| < \sqrt{1-4\mu^2/k^2}$
are both satisfied.  When $|2\alpha -1| < \sqrt{1-4\mu^2/k^2}$,
the amplitude $A_2(k^2)$ has a branch cut (considered as an analytically
continued function of $k^0$) for $(k^0)^2 > \vec{k}^2 + 4\mu^2$.
The discontinuity across the branch cut is related
to the ``cut'' version of the diagram [the second term in 
Eq.~(\ref{eq-dk2})] via the Cutosky rules 
\cite{cutosky:1960a,eden:1966a,thooft:1974a,itzykson:1980a,ramond:1990a,veltman:1994a},
\begin{equation}
\text{tr}_{\text{{\tiny sp}}} [F_{+-}(x,x') F_{-+}(x',x)] 
= -i \int \frac{d^{\,4}k}{(2\pi)^4} e^{-ik(x-x')}
\text{Disc} [ A_2(k^2) ] \theta(k^0).
\label{eq-spm}
\end{equation}
From Eqs.~(\ref{eq-spp}), (\ref{eq-dm}), and (\ref{eq-spm}), it is 
straightforward to obtain an expression for the dissipation kernel,
\begin{equation}
{\mathcal D}_2(x,x') = \frac{3}{4\pi^2} \int \frac{d^{\,4}k}{(2\pi)^4}
e^{-ik(x-x')} \int_0^1 d\alpha E(\alpha;k^2) 
\Biggl[ \text{log}\,\Biggl(\frac{
|E(\alpha;k^2)|}{\Lambda^2}\Biggr) - 
i \pi \theta[-E(\alpha;k^2)] \text{sgn}(k^0) 
\Biggr],
\label{eq-vf1}
\end{equation}
where we have now taken the limit $\epsilon \rightarrow 0_{+}$.
One can verify by inspection that this kernel is real.  However, the
second term in Eq.~(\ref{eq-vf1}) breaks time-reversal invariance and leads to 
dissipative mean field dynamics.  
The one-loop Fourier-transformed mean-field equation is (dropping the
caret from $\hat{\phi}$)
\begin{equation}
\Biggl[ k^2 - m^2 + i k^0 \tilde{\gamma}_2(k) 
- \frac{3 \hbar f^2}{4\pi^2}
\int^1_0 d\alpha E(\alpha; k^2) 
\text{log}\, \Biggl( \frac{|E(\alpha; k^2)|}{\Lambda^2}
\Biggr) \Biggr]
\tilde{\phi}(k) = -\tilde{J}(k),
\label{eq-mfe}
\end{equation}
where $\tilde{\gamma}_2(k)$ is the dissipation function, defined as 
$-i\hbar f^2/k^0$ times
the Fourier transform of the second term in Eq.~(\ref{eq-vf1}),
\begin{equation}
\tilde{\gamma}_2(k) = \frac{\hbar f^2}{k^0} \text{Im}
\,[\tilde{{\mathcal D}}_2(k)] =
\frac{\hbar f^2}{8\pi} \frac{k^2}{|k^0|} \left(1-\frac{4\mu^2}{k^2}
\right)^{3/2} \theta(k^2 - 4\mu^2).
\label{eq-ddf} 
\end{equation}
The one-loop $O(f^2)$ dissipation kernel agrees with previous calculations
of the probability to produce a fermion particle pair of momentum $k$
\cite{abbott:1982a,dolgov:1990a,itzykson:1980a,berera:1995a}.
In Eq.~(\ref{eq-mfe}), $\tilde{J}(k)$ is an external $c$-number source.
The imaginary term $i k^0 \tilde{\gamma}_2 \tilde{\phi}$ in Eq.~(\ref{eq-mfe}) 
breaks time-reversal invariance and acts as a $k$-dependent dissipative force 
in the mean field equation.
The $\theta$ function enforces the energy threshold for the virtual fermion
pair in the one-loop $O(f^2)$ diagram to go on-shell.  The dissipative
mean-field equation (\ref{eq-mfe}) is essentially the linear-response
approximation to the effective inflaton dynamics.  It should be noted
that the dissipation kernel ${\mathcal D}_2(x,x')$ is {\em nonlocal,\/}
in contrast with the local friction-type dissipation assumed in earlier
studies of post-inflation reheating \cite{kolb:1990a}.  However, in the 
limit $\mu^2 \rightarrow 0$, the dissipation kernel does become time-local,
as there is no longer a length scale in the expression for 
${\mathcal D}_2(x,x')$ which could define a time scale for nonlocal
dissipation \cite{calzetta:1988b}.

\subsection{One-loop perturbative effective action at $O(f^4)$}
The $O(f^4)$ term in the one-loop CTP effective action consists of the
``square'' diagram, which is the second term in Fig.~\ref{fig-gam1},
\begin{eqnarray}
\Gamma^{(1)}_4[\phih] = \frac{i}{4} c^{abc} c^{a'b'c'} 
c^{def} c^{d'e'f'}
\int d^4 x d^4 x' d^4 y' d^4 y \biggl[ && \text{tr}_{\text{{\tiny sp}}} 
\bigl[F_{bb'}(x,x') F_{c'f'}(x',y') F_{e'e}(y',y) F_{fc}(y,x)\bigr] 
\nonumber \\ &&
\times \phih_a(x) \phih_{a'}(x') \phih_d (y) 
\phih_{d'}(y')
\biggr].
\end{eqnarray}
Expanding out the contracted CTP indices, we obtain
\begin{eqnarray}
\Gamma^{(1)}_4[\phih] = && \frac{i}{4} \int d^4 x d^4 x' d^4 y' d^4 y 
\biggl[ \nonumber \\
&& \phihp(x) \phihp(x') \phihp(y) \phihp(y') \text{tr}_{\text{{\tiny sp}}} \bigl\{ 
F_{++}(x,x') F_{++}(x',y') F_{++}(y',y) F_{++}(y,x) \bigr\} \nonumber\\
&& + \phihm(x) \phihm(x') \phihm(y) \phihm(y') \text{tr}_{\text{{\tiny sp}}} \bigl\{
F_{--}(x,x') F_{--}(x',y') F_{--}(y',y) F_{--}(y,x) \bigr\} \nonumber \\
&& - 4 \phihp(x) \phihm(x') \phihm(y') \phihm(y) \text{tr}_{\text{{\tiny sp}}} \bigl\{
F_{+-}(x,x') F_{--}(x',y') F_{--}(y',y) F_{-+}(y,x) \bigr\} \nonumber \\
&& - 4 \phihm(x) \phihp(x') \phihp(y') \phihp(y) \text{tr}_{\text{{\tiny sp}}} \bigl\{
F_{-+}(x,x') F_{++}(x',y') F_{++}(y',y) F_{+-}(y,x) \bigr\} \nonumber \\
&& + 4 \phihp(x) \phihp(x') \phihm(y') \phihm(y) \text{tr}_{\text{{\tiny sp}}} \bigl\{
F_{++}(x,x') F_{+-}(x',y') F_{--}(y',y) F_{-+}(y,x) \bigr\} \nonumber \\
&& + 2 \phihp(x) \phihm(x') \phihp(y') \phihm(y) \text{tr}_{\text{{\tiny sp}}} \bigl\{
F_{+-}(x,x') F_{-+}(x',y') F_{+-}(y',y) F_{-+}(y,x) \bigr\} 
\biggr].
\end{eqnarray}
When $\Gamma_4^{(1)}$ is expressed in terms of $\Delta$ and $\Sigma$ 
[defined in Eqs.~(\ref{eq-sdva}) and (\ref{eq-sdvb})], only those
terms with one factor of $\Delta$ and three factors of $\Sigma$ contribute
to the mean field equation of motion.  As a consequence of the unitarity
condition (\ref{eq-uc}),
the sum of terms proportional to four factors of $\Sigma$ must vanish.
Keeping only those terms in the effective action which contribute to the mean
field equation or are quadratic in $\Delta$, we find
\begin{eqnarray}
\Gamma_4^{(1)}[\phih] = \int d^4 x d^4 x' d^4 y' d^4 y \biggl[ &&
\Delta(x) \Sigma(x') \Sigma(y') \Sigma(y) {\mathcal D}_4(x,x',y',y) 
\nonumber \\ && 
+ \frac{i}{2} \Delta(x) \Delta(x') \Sigma(y') \Sigma(y) {\mathcal N}_4
(x,x',y',y) \biggr],
\end{eqnarray}
in terms of a kernel ${\mathcal D}_4(x,x',y',y)$ defined by
\begin{eqnarray}
{\mathcal D}_4(x,x',y',y) = -\text{Im} \, \text{tr}_{\text{{\tiny sp}}}
\biggl[  
&&   F_{++}(x,x') F_{++}(x',y') F_{++}(y',y) F_{++}(y,x) \nonumber \\
&& + F_{++}(x,x') F_{+-}(x',y') F_{--}(y',y) F_{-+}(y,x) \nonumber \\
&& + F_{+-}(x,x') F_{--}(x',y') F_{-+}(y',y) F_{++}(y,x) \nonumber \\
&& + F_{+-}(x,x') F_{-+}(x',y') F_{+-}(y',y) F_{-+}(y,x) \nonumber \\
&& - F_{+-}(x,x') F_{-+}(x',y') F_{++}(y',y) F_{++}(y,x) \nonumber \\
&& - F_{++}(x,x') F_{+-}(x',y') F_{-+}(y',y) F_{++}(y,x) \nonumber \\
&& - F_{++}(x,x') F_{++}(x',y') F_{+-}(y',y) F_{-+}(y,x) \nonumber \\
&& - F_{+-}(x,x') F_{--}(x',y') F_{--}(y',y) F_{-+}(y,x) \label{eq-fod}
\biggr],
\end{eqnarray}
and a ``noise'' kernel ${\mathcal N}_4(x,x',y',y)$ defined by
\begin{eqnarray}
{\mathcal N}_4(x,x',y',y) = \text{Re} \, \text{tr}_{\text{{\tiny sp}}}
\biggl[  
&&   F_{++}(x,x') F_{++}(x',y') F_{++}(y',y) F_{++}(y,x) \nonumber \\
&& + F_{++}(x,x') F_{+-}(x',y') F_{--}(y',y) F_{-+}(y,x) \nonumber \\
&& - F_{+-}(x,x') F_{--}(x',y') F_{-+}(y',y) F_{++}(y,x) \nonumber \\
&& - F_{+-}(x,x') F_{-+}(x',y') F_{+-}(y',y) F_{-+}(y,x) \nonumber \\
&& + F_{+-}(x,x') F_{-+}(x',y') F_{++}(y',y) F_{++}(y,x) \nonumber \\
&& - F_{++}(x,x') F_{+-}(x',y') F_{-+}(y',y) F_{++}(y,x) \nonumber \\
&& - F_{++}(x,x') F_{++}(x',y') F_{+-}(y',y) F_{-+}(y,x) \nonumber \\
&& + F_{+-}(x,x') F_{--}(x',y') F_{--}(y',y) F_{-+}(y,x) \nonumber \\
&& + \{F_{++}(x,y') F_{++}(y',x') F_{++}(x',y) F_{++}(y,x) \nonumber \\
&& - F_{+-}(x,y') F_{--}(y',x') F_{-+}(x',y) F_{++}(y,x) \nonumber \\
&& - F_{++}(x,y') F_{+-}(y',x') F_{--}(x',y) F_{-+}(y,x) \nonumber \\
&& + F_{+-}(x,y') F_{-+}(y',x') F_{+-}(x',y) F_{-+}(y,x) \nonumber \\
&& - F_{+-}(x,y') F_{-+}(y',x') F_{++}(x',y) F_{++}(y,x) \nonumber \\
&& + F_{++}(x,y') F_{+-}(y',x') F_{-+}(x',y) F_{++}(y,x) \nonumber \\
&& - F_{++}(x,y') F_{++}(y',x') F_{+-}(x',y) F_{-+}(y,x) \nonumber \\
&& + F_{+-}(x,y') F_{--}(y',x') F_{--}(x',y) F_{-+}(y,x)\}/2
\biggr],
\end{eqnarray}
The noise kernel ${\mathcal N}_4$ does not contribute to the mean field
equation of motion.  There are, of course, terms in $\Gamma^{(1)}_4[\phih]$
which are higher order in $\Delta$, for example, $O(\Delta^4)$, 
but in passing over to a stochastic equation for $\phih$ in Sec.~\ref{sec-nk},
 we will be assuming that $\Delta$ is small, so that
higher-order terms in powers of $\Delta$ can be ignored.  Such terms will 
in general contribute to non-Gaussian noise,  
which will be studied in an upcoming paper \cite{calzetta:1998b}.  

Let us evaluate the first term of Eq.~(\ref{eq-fod}), 
which consists of only Feynman propagators.  The term is logarithmically
divergent, and as in Sec.~\ref{sec-olpefa}, we use dimensional continuation
and the modified minimal subtraction ($\overline{\text{MS}}$) 
renormalization scheme.  Because we are only interested in 
deriving the dissipative terms in the mean-field equation coming from this
diagram, and because we are assuming $m \gg \mu$, 
we include only the one-loop logarithm.  We find
\begin{eqnarray}
\text{tr}_{\text{{\tiny sp}}} \Bigl[ F_{++}(x,x') && F_{++}(x',y') F_{++}(y',y) F_{++}(y,x) \Bigr]
_{\text{log only}}
\nonumber \\ && = i \int \frac{d^4 k_1}{(2\pi)^4} \frac{d^4 k_2}{(2\pi)^4} 
\frac{d^4 k_3}{(2\pi)^4}  e^{-i [
-(k_1 + k_2 + k_3) \cdot x + k_1 \cdot x' + k_2 \cdot y' + k_3 \cdot y 
]} A_4(k_1, k_2, k_3),
\end{eqnarray}
where the amplitude $A_4(k_1, k_2, k_3)$ is defined by
\begin{equation}
A_4(k_1, k_2, k_3) = -\frac{3}{2\pi^2} \int 
d\alpha_1 d\alpha_2 d\alpha_3 \log \Biggl[ \frac{ E_4
(\alpha_1,\alpha_2,\alpha_3;
k_1,k_2,k_3)}{\Lambda^2} \Biggr],
\end{equation}
in terms of a function $E_4$ defined by
\begin{eqnarray}
E_4(\alpha_1,\alpha_2,\alpha_3;k_1,k_2,k_3) = &&
\Bigl[ (1-\alpha_1)k_1 + (1-\alpha_1 - \alpha_2)k_2
+ (1-\alpha_1 -\alpha_2-\alpha_3)k_3\Bigr]^2 \nonumber 
\\ && - (1-\alpha_1)k_1^2-
(1-\alpha_1-\alpha_2)(k_2^2 + 2k_1\cdot k_2) \nonumber
\\ && 
- (1-\alpha_1-\alpha_2-\alpha_3)(2 k_1 \cdot k_3 + 2 k_2 \cdot k_3 + k_3^2)
+ \mu^2.
\end{eqnarray}
As in Sec.~\ref{sec-olpefa}, the cut diagrams in Eq.~(\ref{eq-fod}) are 
related to the discontinuities in $E_4
(\alpha_1,\alpha_2,\alpha_3;k_1,k_2,k_3)$ 
via the Cutosky rules.  The details are shown in the appendix.
We can then express the dissipation kernel ${\mathcal D}_4(x,x',y',y)$
as a Fourier transform over external momenta,
\begin{equation}
{\mathcal D}_4(x,x',y',y) = 
\int \frac{d^4 k_1}{(2\pi)^4}
\frac{d^4 k_2}{(2\pi)^4} \frac{d^4 k_3}{(2\pi)^4} 
e^{-i[ -(k_1 + k_2 + k_3)\cdot x + k_1 \cdot x' + k_2 \cdot y' + k_3 \cdot y]}
\tilde{{\mathcal D}}_4(k_1,k_2,k_3),
\label{eq-d4ft}
\end{equation}
in terms of a function $\tilde{{\mathcal D}}_4(k_1,k_2,k_3)$ defined by
\begin{eqnarray}
\tilde{{\mathcal D}}_4(k_1,k_2,k_3) = \frac{3}{2\pi} 
\Biggl[ && \frac{1}{\pi}
\int d\alpha_1 d\alpha_2 d\alpha_3 \log \left( \frac{| E_4(\alpha_1,
\alpha_2,\alpha_3; k_1, k_2,k_3) |}{\Lambda^2} \right) \nonumber \\ 
&& +  i\text{sgn}(k_2^0 + k_3^0) h[ (k_2 + k_3)^2 ] +
i\text{sgn}(k_1^0 + k_2^0) h[ (k_1 + k_2)^2 ] \nonumber \\
&& +  i\text{sgn}(k_1^0 + k_2^0 + k_3^0) h[ k_2^2 ] 
+  i\text{sgn}(k_2^0) h[ (k_1 + k_2 + k_3)^2 ] \nonumber \\
&& +  i\text{sgn}(k_3^0) h[ k_1^2 ] +
i\text{sgn}(k_1^0) h[ k_3^2 ] -  i H(k_1,k_2,k_3) \Biggr],
\label{eq-fpdk}
\end{eqnarray}
and the functions $h(s)$ and $H(k_1,k_2,k_3)$ are defined by
\begin{mathletters}
\begin{eqnarray}
&& h(s) = \sqrt{1-\frac{4\mu^2}{s}} \theta(s-4\mu^2), \label{eq-dhf} \\
&& H(k_1,k_2,k_3) = \int_{\alpha_1,\alpha_2,\alpha_3 > 0} \bigl\{  
\theta[ -E_4(\alpha_1,\alpha_2,\alpha_3; k_1,k_2,k_3) ] \nonumber \\
&& \qquad \qquad \qquad \qquad \qquad \qquad 
 \times [ \theta(k_1^0) \theta(-k_2^0) \theta(k_3^0) -
     \theta(-k_1^0) \theta(k_2^0) \theta(-k_3^0) ] \bigr\}.
\end{eqnarray}
\end{mathletters}
Equation~(\ref{eq-fpdk}) leads to the following mean-field equation at
$O(f^4)$,
\begin{eqnarray}
\Biggl[ && k^2 - m^2 + i k^0 \tilde{\gamma}_2(k) -
\frac{3 \hbar f^2}{4\pi^2}
\int^1_0 d\alpha E(\alpha; k^2) 
\text{log}\, \Biggl( \frac{|E(\alpha; k^2)|}{\Lambda^2}
\Biggr) \Biggr]
\tilde{\phi}(k) \nonumber \\
&& 
- \frac{3i\hbar f^4}{2\pi} \int \frac{d^4 q}{(2\pi)^4} \frac{d^4 l}{(2\pi)^4}
\tilde{\phi}(k-q-l) \tilde{\phi}(q) \tilde{\phi}(l) \Biggl[ 
\frac{i}{\pi} \int d\alpha_1 d\alpha_2 d\alpha_3 \log \left( \frac{|E_4
(l+q-k,-q,-l)|}{
\Lambda^2} \right) \nonumber \\
&& \qquad \qquad \qquad\qquad\qquad\qquad \qquad \qquad \qquad \qquad
+ \text{sgn}(q^0 + l^0) h[(q+l)^2] + \text{sgn}(k^0) h(q^2) 
\nonumber \\
&& \qquad \qquad \qquad\qquad\qquad\qquad \qquad \qquad \qquad \qquad
+ \text{sgn}(k^0 - l^0) h[(k-l)^2] + \text{sgn}(q^0) h(k^2) 
\nonumber \\
&& \qquad \qquad \qquad\qquad\qquad\qquad \qquad \qquad \qquad \qquad
+ \text{sgn}(k^0 - q^0 - l^0) h(l^2) \nonumber \\
&& \qquad \qquad \qquad\qquad\qquad\qquad \qquad \qquad \qquad \qquad
+ \text{sgn}(l^0) h[(k-q-l)^2] \nonumber \\
&& \qquad \qquad \qquad\qquad\qquad\qquad \qquad \qquad \qquad \qquad
+ H(l+q-k,-q,-l) 
\Biggr] = -\tilde{J}(k).
\label{eq-og4dt}
\end{eqnarray}
The presence of terms of the form $i\text{sgn}(p^0)h(p^2)$ in Eq.~(\ref{eq-og4dt})
clearly signifies dissipative dynamics.  The $\theta$-function in Eq.~(\ref{eq-dhf})
enforces the energy threshold for the virtual fermion quanta created at a 
particular vertex to go on-shell.
Comparing Eq.~(\ref{eq-og4dt}) and Eq.~(\ref{eq-mfe}), and assuming
spatial homogeneity, we see that the 
$O(f^4)$ dissipation kernel must be taken into account whenever
the condition (\ref{eq-vpt}) fails to hold for the solution $\phi(t)$ to 
Eq.~(\ref{eq-mfe})\@.

At the end of the regime of parametric resonance in chaotic inflaton, i.e.
the ``preheating'' regime, 
the inflaton mean field may oscillate with an amplitude as large as 
$\sim m/g_{\phi\chi}$, where
$g_{\phi\chi}$ is the coupling to another scalar field $\chi$, typically on
the order of $10^{-4}$ \cite{kofman:1997a}.  Condition~(\ref{eq-vpt}) would then
be violated if $f > g_{\phi\chi}$.   In this case it would be necessary, at a 
minimum, to take into account higher order terms (such as ${\mathcal D}_4$)
in the mean-field equation,  until such time as the amplitude $\phih_0(t)$ 
has decreased to the point where Eq.~(\ref{eq-vpt}) is satisfied.

\section{Noise kernel and stochastic inflaton dynamics}
\label{sec-nk}
Although the kernels ${\mathcal N}_2(x,x')$ and ${\mathcal N}_4(x,x',y',y)$
do not contribute to the mean 
field equation, i.e., the equation of motion for $\phih$, they contain
information about stochasticity in a quasi-classical description of the
effective dynamics of the inflaton field 
\cite{hu:1992a,hu:1993a,hu:1993c,hu:1994a,hu:1995a,calzetta:1995a,calzetta:1994a,gleiser:1994a,lombardo:1996a}.  
In this section, we study the effect of stochasticity
on the dynamics of the inflaton mean field, within the perturbative framework
established above.

\subsection{Langevin equation and fluctuation-dissipation relation at
$O(f^2)$}
In this section we show how to obtain a classical stochastic equation 
for the inflaton field from the $O(f^2)$ perturbative CTP
effective action.  From Eq.~(\ref{eq-d1}), it follows that the $O(f^2)$
one-loop perturbative CTP effective action has the form
\begin{equation}
\Gamma[\phih] = {\mathcal S}^{\phi}[\phih] + \int d^4 x d^4 x' \left[
\Sigma(x) \Delta(x') \mu_2(x,x') + 
\frac{i}{2} \Delta(x) \Delta(x') \nu_2(x,x') \right],
\label{eq-olpea2}
\end{equation}
where for simplicity we have defined
\begin{mathletters}
\begin{eqnarray}
\nu_2(x,x') &=& \hbar f^2 {\mathcal N}_2(x,x'), \\
\mu_2(x,x') &=& \hbar f^2 {\mathcal D}_2(x,x').
\end{eqnarray}
\end{mathletters}
In order to extract the stochastic noise arising from the
kernel ${\mathcal N}_2(x,x')$, we use the Gaussian identity \cite{ryder:1985a}
\begin{eqnarray}
\exp\biggl[
-\frac{1}{2\hbar} \int d^{\,4}x d^{\,4}x' \Delta(x) \Delta(x') \nu_2
(x,x') \biggr] 
= N \int D\xi_2 \exp \biggl[  -\frac{1}{2\hbar} && \int d^{\,4}x
d^{\,4}x' \xi_2(x)\nu_2^{-1}(x,x') \xi_2(x') \nonumber \\
&& + \frac{i}{\hbar}\int 
d^{\,4}x \xi_2(x) \Delta(x)\biggr],
\label{eq-gid}
\end{eqnarray}
where $N$ is a normalization factor which does not depend on $\Delta$,
and $\xi_2$ is a $c$-number functional integration variable.
Following \cite{hu:1995a}, we now define a functional
\begin{equation}
P[\xi_2] = N \exp\biggl[-\frac{1}{\hbar} \int d^{\,4}x d^{\,4}x' \xi_2(x) 
\nu_2^{-1}(x,x') \xi_2(x')\biggr],
\label{eq-dou}
\end{equation}
and it follows from Eq.~(\ref{eq-gid}) that $P[\xi_2]$ is normalized in the 
sense of
\begin{equation}
\int D\xi_2 P[\xi_2] = 1.
\end{equation}
Using Eq.~(\ref{eq-gid}), we can rewrite the $O(f^2)$ one-loop CTP effective 
action, Eq.~(\ref{eq-olpea2}), as
\begin{equation}
\Gamma[\phih] = -i \hbar \log \, \int D\xi_2 P[\xi_2] \exp \biggl[ \frac{i}{\hbar}
\Bigl( {\mathcal S}^{\phi}[\phih] + \int d^{\,4}x d^{\,4}x' \Sigma(x)
\Delta(x') \mu_2(x,x') + \int d^{\,4}x \xi_2(x) \Delta(x)
\Bigr) \biggr].
\end{equation}
This suggests defining a new effective action which depends on both
$\xi_2$ and $\phih_{\pm}$ (dropping the carat from $\phih$),
\begin{equation}
\Gamma[\phi,\xi_2] = {\mathcal S}^{\phi}[\phi] + \int d^{\,4}x d^{\,4}x'
\Sigma(x) \Delta(x') \mu_2(x,x') + \int d^{\,4}x \xi_2(x) \Delta(x).
\end{equation}
Let us define a type of ensemble average
\begin{equation}
\langle \langle A \rangle \rangle = \int D\xi_2 P[\xi_2] A(\xi_2),
\label{eq-ntea}
\end{equation}
and note that Eqs.~(\ref{eq-dou}) and (\ref{eq-ntea}) imply that
\begin{mathletters}
\begin{eqnarray}
&& \langle\langle \xi_2(x) \rangle\rangle = 0, \\
&& \langle\langle \xi_2(x)\xi_2(x') \rangle\rangle = \hbar \nu_2(x,x').
\label{eq-nc}
\end{eqnarray}
\end{mathletters}
Clearly then,
\begin{equation}
\left.\left(\frac{\delta}{\delta \phip}\langle\langle \Gamma[\phi,\xi_2] 
\rangle\rangle\right)\right|_{\phip = \phim = \phi} =
\left.\left(\frac{\delta}{\delta \phip}\Gamma[\phi] \right)
\right|_{\phip = \phim = \phi}.
\end{equation}
Taking the variation of $\Gamma[\phi,\xi_2]$ with respect to $\phip$ and
setting $\phip = \phim = \phi$, we obtain (after a Fourier 
transform)
\begin{equation}
\Biggl[ k^2 - m^2 + i k^0 \tilde{\gamma}_2(k) -
\frac{3 \hbar f^2}{4\pi^2}
\int^1_0 d\alpha E(\alpha;k^2) \text{log}\, 
\Biggl( \frac{|E(\alpha;k^2)|}{\Lambda^2}
\Biggr)\Biggr] \tilde{\phi}(k) + \tilde{\xi}_2(k) = -\tilde{J}(k),
\label{eq-le}
\end{equation}
where $\tilde{\xi}_2(k)$ is defined by
\begin{equation}
\tilde{\xi}_2(k) = \int d^{\,4}x e^{ikx} \xi_2(x).
\end{equation}
We now interpret Eq.~(\ref{eq-le}) as a Langevin equation with 
stochastic force $\xi_2$.  The inflaton Fourier mode $\tilde{\phi}$ 
appearing in Eq.~(\ref{eq-le}) should be viewed as a $c$-number 
stochastic variable, and the presence of the 
stochastic force $\xi_2$ indicates 
spontaneous breaking of spatial translation invariance by a Gaussian
(but not white) noise source $\xi_2$ \cite{hu:1994a}.  
Moreover, this stochastic equation obeys a
zero-temperature fluctuation-dissipation relation, as we now show.
First, let us calculate the one-loop $O(f^2)$ noise kernel, 
${\mathcal N}_2(x,x')$ [defined in Eq.~(\ref{eq-dn2}) above], using dimensional
regularization and modified minimal subtraction,
\begin{equation}
\nu_2(x,x') = \frac{\hbar f^2}{8\pi} \int \frac{d^{\,4}k}{(2\pi)^4}
e^{-ik(x-x')} \tilde{\nu}_2(k),
\end{equation}
in terms of the Fourier-transformed noise kernel
\begin{equation}
\tilde{\nu}_2(k) = \frac{\hbar f^2}{8\pi} k^2 \Bigl( 1-\frac{4\mu^2}{
k^2}\Bigr)^{\frac{3}{2}} \theta(k^2 - 4\mu^2).
\label{eq-dnk}
\end{equation}
The noise kernel $\nu_2(x,x')$ is colored; colored noise has been observed in 
other interacting field theories \cite{hu:1993a,hu:1994a,hu:1995a}.
By inspection of Eqs.~(\ref{eq-ddf}) and (\ref{eq-dnk}), it follows that
\begin{equation}
|k^0| \tilde{\gamma}_2(k) = \tilde{\nu}_2(k),
\label{eq-fdr1}
\end{equation}
which leads to the zero-temperature  
fluctuation-dissipation relation \cite{hu:1995a},
\begin{equation}
\nu_2(t,\vec{k}) = \int_{-\infty}^{\infty} dt' K(t-t') 
\gamma_2(t',\vec{k}),
\end{equation}
in terms of the distribution-valued kernel $K(t)$ defined by
\begin{equation}
K(t) = \int_0^{\infty} \frac{d\omega}{\pi} \omega \cos (\omega t),
\end{equation}
and the spatially Fourier-transformed dissipation function and noise kernel,
\begin{mathletters}
\begin{eqnarray}
&& \nu_2(t,\vec{k}) = \int^{\infty}_{-\infty}
 \frac{d k^0}{2\pi} e^{-ik^0 t} \tilde{\nu}_2(k), \\
&& \gamma_2(t,\vec{k}) = \int^{\infty}_{-\infty} 
\frac{d k^0}{2\pi} e^{-ik^0 t} \tilde{\gamma}_2(k).
\end{eqnarray}
\end{mathletters}
This shows the physical significance of the noise kernel $\nu_2(x,x')$
in an effective description of the dynamics of the scalar mean field.  

\subsection{Langevin equation and fluctuation-dissipation relation
at $O(f^4)$}
In this section we consider the $O(f^4)$ one-loop noise kernel,
${\mathcal N}_4$.  The non-normal-threshold
singularities in $A_4$ lead to a noise kernel which depends on $\Sigma$, 
which is known to lead to ambiguities in the resulting Langevin equation
\cite{risken:1989a,son:1997a}.  The meaning and interpretation of
the non-normal-threshold parts of ${\mathcal N}_4$ and ${\mathcal D}_4$
will be the subject
of a future study \cite{calzetta:1998b}.  Here, we focus on the effect of the 
{\em normal-threshold\/} singularities of $A_4$, which for the noise kernel, 
${\mathcal N}_4$, contribute a term
\begin{equation}
\frac{i}{2} 
\int d^4x d^4 x' \Delta(x) \Delta(x') \Sigma(x) \Sigma(x') \nu_4(x,x')
\end{equation}
to the CTP effective action, where the kernel $\nu_4(x,x')$ is defined by
\begin{equation}
\nu_4(x,x') = -\frac{3\hbar f^4}{\pi} \int \frac{d^4 q}{(2\pi)^4}
e^{-i q\cdot(x-x')} h(q^2),
\end{equation}
and the function $h(s)$ was defined in Eq.~(\ref{eq-dhf}) above.  The 
normal-threshold singularities of the dissipation kernel, ${\mathcal D}_4$,
[the second and third terms of Eq.~(\ref{eq-fpdk})], lead to the following 
contribution to the CTP effective action,
\begin{equation}
\int d^4 x d^4 x' \Delta(x) \Sigma(x) \left[ \Sigma(x') \right]^2 \mu_4(x,x'),
\end{equation}
where the kernel $\mu_4(x,x')$ is defined by
\begin{equation}
\mu_4(x,x') = -\frac{3 i \hbar f^4}{\pi} \int \frac{d^4 q}{(2\pi)^4} 
e^{-i q\cdot(x'-x)} \text{sgn}(q^0) h(q^2).
\end{equation}
\begin{mathletters}
With the definitions
\begin{eqnarray}
&& \mu_4(x,x') = i \int \frac{d^4 q}{(2\pi)^4} e^{-i q\cdot(x'-x)}
q^0 \tilde{\gamma}_4(q), \\
&& \nu_4(x,x') = \int \frac{d^4 q}{(2\pi)^4} e^{-i q\cdot(x'-x)}
\tilde{\nu}_4(q),
\end{eqnarray}
\end{mathletters}
it follows immediately that the normal-threshold parts of ${\mathcal D}_4$
and ${\mathcal N}_4$ obey a fluctuation-dissipation
relation identical in form to Eq.~(\ref{eq-fdr1}),
\begin{equation}
|q^0| \tilde{\gamma}_4(q) = \tilde{\nu}_4(q).
\end{equation}
Making use of Eq.~(\ref{eq-gid}), the $O(f^4)$ effective action (including
only normal-threshold contributions) can be written in the form
\begin{eqnarray}
\Gamma[\phi,\xi_2,\xi_4] = && {\mathcal S}^{\phi}[\phi] 
+ \int d^4 x d^4 x' \Delta(x) \Sigma(x') \mu_2(x,x') 
+ \int d^4 x \xi_2(x) \Delta(x)  \nonumber \\ &&
+ \int d^4 x d^4 x' \Delta(x) \Sigma(x) \left[\Sigma(x')\right]^2\mu_4(x,x') 
+ \int d^4 x \xi_4(x) \Delta(x) \Sigma(x),
\label{eq-nea}
\end{eqnarray}
where the stochastic noise source $\xi_4$ satisfies the conditions
\begin{mathletters}
\begin{eqnarray}
&& \langle\langle \xi_4(x) \rangle\rangle = 0 \\ 
&& \langle\langle \xi_4(x) \xi_4(x') \rangle\rangle = \hbar \nu_4(x,x').
\end{eqnarray}
\end{mathletters}
Taking the functional derivative of Eq.~(\ref{eq-nea}) and making the
usual identification, we obtain a Langevin equation with an additive
noise $\xi_2$ and a multiplicative noise $\xi_4$,
\begin{eqnarray}
\biggl[ && q^2 - m^2 + i q^0 \tilde{\gamma}_2(q) - \frac{3 \hbar f^2}{4\pi^2}
\int_0^1 d\alpha E(\alpha;q^2) \log \left( \frac{|E(\alpha;q^2)|}{\Lambda^2}
\right) \biggr] \tilde{\phi}(q) \nonumber \\ && 
+ \int \frac{d^4 k}{(2\pi)^4} 
\frac{d^4 l}{(2\pi)^4} \tilde{\phi}(q-l) \tilde{\phi}(k)
\tilde{\phi}(l-k)  \left[ i l^0 \tilde{\gamma}_4(l) - \frac{3\hbar f^4}{2\pi^2}
\int_0^{1} d^3 \alpha \log\left( \frac{|E_4(
\alpha_1,\alpha_2,\alpha_3; l-q,-k,l-k)|}{\Lambda^2}\right)\right] \nonumber \\
&& \qquad\qquad\qquad\qquad\qquad\qquad\qquad\qquad
= -\tilde{\xi}_2(q) -\tilde{J}(q) -\int \frac{d^4 k}{(2\pi)^4} \tilde{\xi}_4
(q-k) \tilde{\phi}(k),
\end{eqnarray}
where $d^3 \alpha = d\alpha_1 d\alpha_2 d\alpha_3$.
The stochastic force $\xi_4$ is clearly seen to contribute multiplicatively
to the Langevin equation for $\phih$.

\subsection{Homogeneous mean field dynamics at $O(f^2)$}
To make connection with post-inflationary reheating, it is customary to
assume that the mean field $\phi$ is spatially homogeneous 
\cite{kofman:1996a,kofman:1997a,shtanov:1995a}.  In this case,
the Langevin equation~(\ref{eq-le}) takes the form
\begin{equation}
\left[\omega^2 - m^2 + i \omega \beta(\omega) + \eta(\omega) 
\right] \tilde{\phi}(\omega) + \tilde{\xi}_2(\omega) = -\tilde{J}(\omega),
\label{eq-hle}
\end{equation}
where we have defined
\begin{mathletters}
\begin{eqnarray}
&& \beta(\omega) = \frac{\hbar f^2
}{8\pi}\frac{\omega^2}{|\omega|}\left(
1-\frac{4\mu^2}{\omega^2} \right)^{3/2}\theta(\omega^2 - 4\mu^2), \\
&& \eta(\omega) = -\frac{3\hbar f^2}{4\pi^2}\int_0^1 d\alpha E(\alpha;\omega^2)
\log \left( \frac{|E(\alpha;\omega^2)|}{\Lambda^2} \right). 
\end{eqnarray}
\end{mathletters}
The total energy dissipated to the fermion field over the history of
the dynamical evolution of the mean field is given [at $O(f^2)$] by
\begin{equation}
{\mathcal E} = -\int^{\infty}_{-\infty} dt F_v(t) \frac{d \phi(t)}{dt},
\end{equation}
where the friction force $F_v(t)$ is the Fourier transform of $i\beta
(\omega) \omega \tilde{\phi}(\omega)$.  
After a bit of Fourier algebra, we obtain an expression for the 
ensemble-averaged, total dissipated energy,
\begin{equation}
\langle\langle{\mathcal E}\rangle\rangle = \frac{\hbar f^2
}{8\pi^2} \int_{2\mu}^{\infty} d\omega
\omega^3 \left( 1- \frac{4 \mu^2}{\omega^2} \right)^{3/2} \frac{
|\tilde{J}
(\omega)|^2}{\left[ \omega^2 - m^2 + \eta(\omega) 
\right]^2 + \omega^2 \beta(\omega)^2}.
\end{equation}
It is straightforward to compute the variance in the total dissipated energy.
Making use of Eq.~(\ref{eq-nc}), we find
\begin{equation}
|\langle\langle {\mathcal E}^2 \rangle\rangle - \langle\langle {\mathcal E}
\rangle\rangle^2| = \frac{\hbar^4 f^6}{256 \pi^6}\int_{2\mu}^{\infty}
d\omega \omega^6 I(\omega)
 \left( 1-\frac{4\mu^2}{\omega^2} \right)^{3}
\frac{|\tilde{J}(\omega)|^2}{\left[ (\omega^2 - m^2 + 
\eta(\omega))^2 + \omega^2 \beta(\omega)^2\right]^2},
\end{equation}
where the function $I(\omega)$ is defined by
\begin{equation}
I(\omega) = \int_0^{\sqrt{\omega^2 - 4\mu^2}} dk k^2 (\omega^2 - k^2) \left(
1-\frac{4\mu^2}{\omega^2 - k^2}\right)^{3/2}.
\end{equation}
Following \cite{abbott:1982a}, we assume that the inflaton field
is held fixed via an external, constant
 $c$-number source $J$ for $t < 0$, and that
the source is removed for $t \geq 0$,
\begin{equation}
J(t) = J \theta(-t).
\label{eq-jt}
\end{equation}
Setting $\hbar = 1$, assuming that $m \gg \mu$, and expanding to lowest order 
in $f$, we obtain for the ensemble averaged dissipated energy,
\begin{equation}
\langle\langle {\mathcal E} \rangle\rangle = \frac{f^2 J^2}{16\pi^2 m^2}.
\label{eq-eade}
\end{equation}
Let us now compute the variance in the total dissipated energy.
Performing a regularization via dimensional continuation, we obtain 
\begin{equation}
|\langle\langle {\mathcal E}^2 \rangle\rangle - 
\langle\langle {\mathcal E} \rangle\rangle^2| =
\frac{f^6 J^2 m^2}{960 \pi^6} \delta^2,
\label{eq-vtde}
\end{equation}
where $\delta$ is a constant of order unity defined by
$\delta^2 = | 119/60 - \gamma_{\text{{\tiny EM}}} -
\log (4\pi m^2/\Lambda^2)|$, and 
$\gamma_{\text{{\tiny EM}}}$ is the Euler-Mascheroni
constant, $\approx 0.5772$.
Taking the ratio of the square root of Eq.~(\ref{eq-vtde}) and 
Eq.~(\ref{eq-eade}), we obtain the relative strength of the rms fluctuations
in the total dissipated energy density, ${\mathcal E}_{\text{{\tiny rms}}}$, 
\begin{equation}
\frac{{\mathcal E}_{\text{{\tiny rms}}}}{
\langle\langle {\mathcal E}\rangle\rangle} \equiv
\frac{\sqrt{|\langle\langle {\mathcal E}^2 \rangle\rangle -
\langle\langle {\mathcal E} \rangle\rangle^2|}}{\langle\langle
{\mathcal E}\rangle\rangle} 
= \frac{2 f m^3 \delta}{\sqrt{15} \pi J}.
\end{equation}
The parameter $J$ is related to the initial inflaton amplitude 
$\phih_0(t_0)$ by
$J = \phih_0(t_0) m^2 /2$, which leads to
\begin{equation}
\frac{{\mathcal E}_{\text{{\tiny rms}}}}{
\langle\langle {\mathcal E} \rangle\rangle} = 
\frac{4 f m \delta}{\sqrt{15} \pi \phih_0(t_0)} \simeq 0.390 
\frac{m f}{\phih_0(t_0)}.
\label{eq-dee}
\end{equation}
The fundamental assumption which justified the perturbative expansion in $f$,
Eq.~(\ref{eq-vpt}), is seen to be independent of 
Eq.~(\ref{eq-dee}).  Therefore, the ratio 
${\mathcal E}_{\text{{\tiny rms}}} /
{\mathcal E}$ is not required to be small by consistency with 
perturbation theory.  As the initial inflaton amplitude $\phi_0$ is
made larger, the relative strength of the rms fluctuations of ${\mathcal E}$
is seen to decrease, in accordance with the correspondence principle.
It has been shown that the fluctuations in the total dissipated energy
density are related to the fluctuations in the occupation numbers of modes
\cite{calzetta:1994a}.
 
Let us now examine whether the rms fluctuations in the total dissipated
energy, as given by Eq.~(\ref{eq-dee}), is significant, given a reasonable
value for the inflaton amplitude at the end of the preheating regime
(the period of parametric resonance-induced particle production).
In chaotic inflaton with a scalar field $\chi$ coupled to the inflaton field
via a coupling constant $g_{\phi\chi}$, the typical inflaton amplitude at
the end of the preheating regime is on the order of $m/g_{\phi\chi}$ 
\cite{kofman:1997a}.  In this case, we would find
${\mathcal E}_{\text{{\tiny rms}}}/\langle\langle {\mathcal E}\rangle\rangle
\simeq f g_{\phi\chi},$
from which it is clear that fluctuations in the {\em total\/} dissipated energy
are not significant relative to the ensemble-averaged total dissipated energy,
and therefore should not appreciably affect the reheating temperature.  
However, in {\em new inflation\/} scenarios where
the inflaton amplitude $\phih_0$ can be on the order of $m$ at the onset of
reheating, the ratio $m/\phih_0$ can be of order unity \cite{kolb:1990a}.  
In this case, the ratio ${\mathcal E}_{\text{{\tiny rms}}}/
\langle\langle{\mathcal E}
\rangle\rangle \simeq f$, which may not be a negligible effect.

Although as shown above, stochasticity does not dramatically affect the
total energy dissipated via fermion production in chaotic inflation, we may
inquire whether the noise term in the Langevin equation for the inflaton 
zero-mode, Eq.~(\ref{eq-hle}), may nonetheless be non-negligible during the 
reheating period.   Let us compute the rms fluctuations in the inflaton 
zero-mode, $\phih_{\text{{\tiny rms}}}$.  
Following methods described above, we find that the rms fluctuations of
the inflaton zero mode are given, to $O(f^2)$, by
\begin{equation}
\phih_{\text{{\tiny rms}}} = \frac{f}{\pi} \frac{m}{\sqrt{60\pi}} \sigma,
\label{eq-frms}
\end{equation}
where $\sigma^2 = |61/30 - \gamma_{\text{{\tiny EM}}} - \log (4\pi m^2/
\Lambda^2)|$.  Equation~(\ref{eq-frms}) is seen to be independent of the 
inflaton zero-mode amplitude $\phih_0$.

In order to determine the relative importance of fluctuations in the 
inflaton zero-mode amplitude $\phih_0$ during and at the end of the
reheating period, we must introduce curved spacetime arguments.  This is
because the end of the reheating period is determined by the time 
$t_{\text{{\tiny end}}}$ at which
the Hubble constant becomes of the order of $3\beta(m)$
for the case of $\lambda=0$ being discussed in this section 
\cite{kolb:1990a,shtanov:1995a}. 
Starting with the semiclassical Einstein equation (\ref{eq-see}) for spatially
flat Friedmann-Robertson-Walker (FRW) cosmology, setting 
$b=c=\Lambda_c=0$ (following arguments similar to those of Sec.~III~D of
Ref.~\cite{ramsey:1997b}), retaining only the inflaton zero mode as the
dynamical degree of freedom (for consistency with FRW), and retaining
both the $\psi$ field energy density $\rho_{\psi}$ and the 
{\em classical, stochastic\/} energy density of the inflaton zero mode, we have
\begin{equation}
H^2 = \frac{\dot{a}^2}{a^2} = \frac{8\pi}{3\Mpl^2} \left(
 \rho(\overline{\phih^2}) + \rho_{\psi} \right),
\label{eq-fre}
\end{equation}
where $a$ is the scale factor, the dot denotes a derivative with respect to
cosmic time, and $\rho(\overline{\phih^2})$ is the energy density
as a function of the time-average (over one period of oscillation)
of $\phih^2$, which is given by the virial theorem,
\begin{equation}
\overline{\rho(\phih)} \simeq m^2 \overline{\phih^2} = \frac{1}{2} m^2 
(\phih_0)^2.
\end{equation}
Making use of Eqs.~(\ref{eq-hle}), (\ref{eq-jt}), and (\ref{eq-fre}),
we obtain an approximate expression for the (ensemble-averaged)
energy density of the inflaton zero-mode at the end of the reheating
period\footnote{
We wish to emphasize, however, that this expression does {\em not\/} take
into account the regime of nonperturbative dynamics discussed in
Sec.~\ref{sec-necgidcst}, and therefore should not be expected to yield
a correct reheating temperature in a realistic inflationary scenario.  
However, it suffices for the present discussion of rms fluctuations
of the inflaton amplitude, where we assume an idealized case similar to
Eq.~(106) of Ref.~\cite{shtanov:1995a}.}
 (to lowest order in $f$),
\begin{equation}
\rho(t_{\text{{\tiny end}}}) \simeq \frac{3 f^4 \Mpl^2 m^2}{(8\pi)^3 e},
\label{eq-endrh}
\end{equation}
where $e$ is the base of the natural logarithm.  Note that this expression
is independent of the initial inflaton amplitude \cite{kofman:1997a}.
Equation~(\ref{eq-endrh}) allows us to solve 
for the value of $\phih_0$ at the end of reheating.  We find
\begin{equation}
\phih_0(t_{\text{{\tiny end}}}) \simeq \frac{\sqrt{6/e} \Mpl f^2}{
(8\pi)^{3/2}}.
\label{eq-minp}
\end{equation}
The rms fluctuations in the inflaton zero mode, $\phih_{\text{{\tiny rms}}}$,
can only play a role in the inflaton zero-mode dynamics during reheating
if the ratio $\phih_{\text{{\tiny rms}}} 
/ \phih_0$ is not small relative to higher-order
[e.g., $O(f^4)$] processes which we are neglecting.  In light of the minimum
inflaton zero-mode amplitude attained during reheating, Eq.~(\ref{eq-minp}),
we find that the ratio of fluctuations in the inflaton zero-mode to the
zero-mode amplitude is given by
\begin{equation}
\frac{\phih_{\text{{\tiny rms}}}}{\phih_0(t_{\text{{\tiny end}}})} \simeq
\frac{8 \sigma m}{f \Mpl} \sqrt{\frac{e}{45}} \simeq 2.37 \frac{m}{f \Mpl}.
\label{eq-vizm}
\end{equation}
We estimate the ratio of the mean-squared inflaton amplitude 
fluctuations to the shift in the inflaton mass to be
\begin{equation}
\left|\frac{\hat{\phi}^2_{\mbox{\tiny rms}}}{\eta(m)}\right| \simeq 0.01.
\end{equation}
If, prior to the end of reheating at $t_{\text{{\tiny end}}}$,
 $\phih_{\text{{\tiny rms}}}/\phih_0(t)$ becomes larger than higher-order
terms which are neglected in our perturbative expansion, then
fluctuations in the inflaton zero mode are a non-negligible effect.
This will happen when $\phih_{\text{{\tiny rms}}}/\phih_0(t_{\text{{\tiny 
end}}})$, given by Eq.~(\ref{eq-vizm}), is not $\ll 1$.
This shows that
stochasticity must be taken into account in the dynamics of the
inflaton zero mode, during the late stages of the reheating 
period.

\section{Conclusions}
In this paper, we present the results of a study of (unbroken-symmetry) 
inflaton dynamics
during the late stages of reheating, which is dominated by fermion 
particle production to a light spinor field coupled to the inflaton
field via a Yukawa coupling.  We derived coupled nonperturbative 
equations for the inflaton mean field and two-point function, in a 
general curved spacetime, and showed that, in addition to the
dissipative mean-field equation, the gap equation for the two-point
function is also dissipative, due to fermion particle production.
Simultaneous evolution of the inflaton mean-field and two-point function
is necessary for correctly following the inflaton dynamics after the
end of the preheating period, because the large value of the variance
invalidates use of the ordinary perturbative, 1PI effective action.

We also derived the dissipation and noise kernels for the small-amplitude
dynamics of the inflaton field, valid in the late stages of reheating
when the inflaton mean-field amplitude is very small.  The $O(f^2)$ noise
and dissipation kernels, as well as the normal-threshold parts of the
$O(f^4)$ noise and dissipation kernels, are shown to obey a zero-temperature 
fluctuation-dissipation relation.  With the noise and dissipation kernels,
a Langevin equation for the inflaton zero mode is derived, and it is shown
that the noise leads to a variance for the inflaton amplitude which
is non-negligible before the end of reheating.

\label{sec-conc}

%%%%%%%%%%%%%%%%%%%%%%%%%%%%%%%%%%%%%%%%%%%%%%%%%%%%%%%%%%%%%%%%%%%%%%%%%%%%
\section{Acknowledgments}
This work is supported in part by NSF Grant No.\ 
PHY94-21849.  Part of this work was carried out at the Institute for 
Advanced Study, Princeton, where B.L.H. was a Dyson Visiting Professor.
We enjoyed the hospitality of Los Alamos National Laboratory during the
Santa Fe workshop ``Nonequilibrium Phase Transitions'' sponsored by
the Center for Nonlinear Studies, and organized by Dr.~E.~Mottola. We thank
Dr.~E.~Calzetta for discussions.  S.A.R. wishes to thank the University of
Buenos Aires for hospitality, and G.~Stephens for helpful discussions.

\appendix
\section{Discontinuities of the square diagram}
\label{sec-cutdiag}
In this appendix, the seven terms of Eq.~(\ref{eq-fod}) involving cut 
propagators are explicitly evaluated using the Cutosky rules.
The second and third terms of Eq.~(\ref{eq-fod}) correspond to normal-threshold
singularities in the $t$ and $s$ channels, and are given by
\begin{eqnarray} 
\int \frac{d^4 q}{(2\pi)^4}
\text{tr}_{\text{{\tiny sp}}}\Bigl[ F_{++}(q) F_{+-}(q+k_1) && 
F_{--}(q+k_1+k_2) F_{-+}(q+k_1+k_2+k_3) \Bigr] 
\nonumber \\ && =
 -i \text{Disc}[A_4(k_1,k_2,k_3)_{|\alpha_1 = \alpha_3 = 0}
] \theta(k_2^0 + k_3^0) 
\end{eqnarray}
and
\begin{eqnarray} 
\int \frac{d^4 q}{(2\pi)^4}
\text{tr}_{\text{{\tiny sp}}}\Bigl[ F_{+-}(q) F_{--}(q+k_1) && 
F_{-+}(q+k_1+k_2) F_{++}(q+k_1+k_2+k_3) \Bigr] 
\nonumber \\ && =
 -i \text{Disc}[A_4(k_1,k_2,k_3)_{|\alpha_2=0; \;\; 
\alpha_1 + \alpha_3 = 1}] \theta(k_1^0 + k_2^0),
\end{eqnarray}
respectively.  The third term in Eq.~(\ref{eq-fod}) corresponds to the 
leading-order singularity of the square diagram \cite{eden:1966a} (i.e.,
the solution of the Landau equations in which $\alpha_1, \alpha_2, \alpha_3$
are all nonzero),
\begin{eqnarray}
\int \frac{d^4 q}{(2\pi)^4}
\text{tr}_{\text{{\tiny sp}}}\Bigl[ F_{+-}(q) F_{-+}(q+k_1) && 
F_{+-}(q+k_1+k_2) F_{-+}(q+k_1+k_2+k_3) \Bigr] 
\nonumber \\ && =
 i \text{Disc}[A_4(k_1,k_2,k_3)_{|\alpha_1,\alpha_2,\alpha_3 > 0}
] \theta(k_1^0) \theta(-k_2^0) \theta(k_3^0).
\end{eqnarray}
The last four terms in Eq.~(\ref{eq-fod}) correspond to the four remaining
twice-contracted singularities, and are given by
\begin{eqnarray}
\int \frac{d^4 q}{(2\pi)^4}
\text{tr}_{\text{{\tiny sp}}}\Bigl[ F_{++}(q) F_{+-}(q+k_1) && 
F_{-+}(q+k_1+k_2) F_{++}(q+k_1+k_2+k_3) \Bigr] 
\nonumber \\ && =
 i \text{Disc}[A_4(k_1,k_2,k_3)_{|\alpha_2 = \alpha_3 = 0}
] \theta(k_2^0),
\end{eqnarray}
\begin{eqnarray}
\int \frac{d^4 q}{(2\pi)^4}
\text{tr}_{\text{{\tiny sp}}}\Bigl[ F_{++}(q) F_{++}(q+k_1) && 
F_{+-}(q+k_1+k_2) F_{-+}(q+k_1+k_2+k_3) \Bigr] 
\nonumber \\ && =
 i \text{Disc}[A_4(k_1,k_2,k_3)_{|\alpha_3 = 0; 
\;\; \alpha_1 + \alpha_2 = 1}] \theta(k_3^0),
\end{eqnarray}
\begin{eqnarray}
\int \frac{d^4 q}{(2\pi)^4}
\text{tr}_{\text{{\tiny sp}}}\Bigl[ F_{+-}(q) F_{-+}(q+k_1) && 
F_{++}(q+k_1+k_2) F_{++}(q+k_1+k_2+k_3) \Bigr] 
\nonumber \\ && =
 i \text{Disc}[A_4(k_1,k_2,k_3)_{|\alpha_1 = \alpha_2 = 0}
] \theta(k_1^0),
\end{eqnarray}
\begin{eqnarray}
\int \frac{d^4 q}{(2\pi)^4}
\text{tr}_{\text{{\tiny sp}}}\Bigl[ F_{+-}(q) F_{--}(q+k_1) && 
F_{--}(q+k_1+k_2) F_{-+}(q+k_1+k_2+k_3) \Bigr] 
\nonumber \\ && =
 i \text{Disc}[A_4(k_1,k_2,k_3)_{|\alpha_1 = 0; \;\; 
\alpha_2 + \alpha_3 = 1}] \theta(k_1^0 + k_2^0 + k_3^0).
\end{eqnarray}

%%%%%%%%%%%%%%%%%%%%%%%%%%%%%%%%%%%%%%%%%%%%%%%%%%%%%%%%%%%%%%%%%%%%%%%%%%%%
%                                                                          %
% The Bibliography (from file "master.bib")                                %
%                                                                          %
%%%%%%%%%%%%%%%%%%%%%%%%%%%%%%%%%%%%%%%%%%%%%%%%%%%%%%%%%%%%%%%%%%%%%%%%%%%%
%\bibliographystyle{prsty}
%\bibliography{master}

%%%%%%%%%%%%%%%%%%%%%%%%%%%%%%%%%%%%%%%%%%%%%%%%%%%%%%%%%%%%%%%%%%%%%%%%%%%%
%                                                                          %
% Figures                                                                  %
%                                                                          %
%%%%%%%%%%%%%%%%%%%%%%%%%%%%%%%%%%%%%%%%%%%%%%%%%%%%%%%%%%%%%%%%%%%%%%%%%%%%
\newpage

\end{document}